\documentclass{pasj00}

\begin{document}

\SetRunningHead{Y. Takeda et al.}{Spectrum variability of $\xi$ Boo A}
\Received{2019/12/06}
\Accepted{2020/01/13}

\title{Spectrum variability of the active solar-type star $\xi$~Bootis~A
}

%

\author{
Yoichi \textsc{Takeda,}\altaffilmark{1,2}
Satoshi \textsc{Honda,}\altaffilmark{3}
Hikaru \textsc{Taguchi,}\altaffilmark{4} and
Osamu \textsc{Hashimoto}\altaffilmark{4}
}

\altaffiltext{1}{National Astronomical Observatory, 2-21-1 Osawa, 
Mitaka, Tokyo 181-8588, Japan}
\email{takeda.yoichi@nao.ac.jp}
\altaffiltext{2}{SOKENDAI, The Graduate University for Advanced Studies, 
2-21-1 Osawa, Mitaka, Tokyo 181-8588, Japan}
\altaffiltext{3}{Nishi-Harima Astronomical Observatory, Center for Astronomy,\\
University of Hyogo, 407-2 Nishigaichi, Sayo-cho, Sayo, Hyogo 679-5313, Japan}
\altaffiltext{4}{Gunma Astronomical Observatory, 6860-86 Nakayama, 
Takayama, Agatsuma, Gunma 377-0702, Japan}


\maketitle

\begin{abstract}
An extensive spectroscopic study on $\xi$~Boo~A (chromospherically active solar-type 
star) was conducted based on the spectra obtained in 2008 December though 2010 May, 
with an aim to detect any spectrum variability and to understand its physical origin. 
For each spectrum, the atmospheric parameters were spectroscopically determined based on 
Fe lines, and the equivalent widths (along with the line-broadening parameters) of 
selected 99 lines were measured. We could detect meaningful small fluctuations 
in the equivalent widths of medium-strength lines. This variation was found 
to correlate with the effective temperature ($T_{\rm eff}$) consistently with the 
$T$-sensitivity of each line, which indicates that the difference in the mean 
temperature averaged over the disk of inhomogeneous condition is mainly responsible 
for this variability. It was also found that the macrobroadening widths of 
medium-strength lines and the equivalent widths dispersion of saturated lines
tend to increase with the effective Land\'{e} factor, suggesting an influence of 
magnetic field. Our power spectrum analysis applied to the time-sequence data of 
V~{\sc i}/Fe~{\sc ii} line-strength ratio and $T_{\rm eff}$ could not confirm the 
6.4~d period reported by previous studies. We suspect that surface inhomogeneities of 
$\xi$~Boo~A at the time of our observations were not so much simple (such as single
star patch) as rather complex (e.g., intricate aggregate of spots and faculae).  
\end{abstract}

%


\section{Introduction}

The solar-type star $\xi$~Boo~A (= HR~5544 = HD~131156A; $V = 4.68$~mag, G7~Ve),  
the primary star of the  visual binary system (151~yr period) in companion with 
$\xi$~Boo~B ($V = 6.82$~mag; K5~Ve), has attracted interest of stellar 
astrophysicists because of its especially high activity, which is manifested by 
strong chromospheric emissions in its spectrum. 
While a number of investigations have been published so far regarding this star,
it was not until Toner and Gray's (1988) pioneering paper that a comprehensive study 
was tried toward clarifying the nature of inhomogeneous surface structure. 
The conclusions of their spectral line-profile analysis based on long-term 
monitoring observations in 1984--1987 (e.g., a large star patch covering 
$\sim 10$\% of the disk which causes the rotational modulation with a period of 
6.43~d, the patch being cooler by $\sim 200$~K with larger velocity dispersion 
than the surrounding though appreciable sign of magnetic field is not seen, 
no significant changes were observed over the 4-year period, etc.) were somewhat 
surprising because of the considerable difference as compared to the solar case. 
Toner and LaBonte (1991) then proposed another explanation invoking Evershed flow in
the penumbra  of star patch (instead of the increase in the velocity dispersion).
Gray et al. (1996) further investigated  the correlation between the long-term variation
of stellar activity during the 1984--1993 period  (based on Ca~{\sc ii} HK line 
core emission) and the properties of star patch, and found a phase delay 
by $\sim$~1.5--2 years between them. 

Meanwhile, notable progress regarding the nature of magnetic fields in $\xi$~Boo~A
has been made in this century. Although its detection was first reported by 
Robinson, Worden, and Harvey's (1980) early observation with Zeeman analyzer, 
the recent results were derived with the help of the more efficient instrument 
along with the modern reduction technique:
Petit et al. (2005) discussed the geometry of large-scale magnetic field based on
their spectropolarimetric observations over 40 nights in 2003.
Morgenthaler et al. (2012) analyzed 76 spectra observed in 2007--2011 by 
spectropolarimeter, and investigated the long-term temporal evolution of the
magnetic field as well as its relation to activity indicators.
Very recently, Cotton et al. (2019) confirmed the existence of rotationally 
modulated polarization (with the rotation period of 6.4~d derived by Toner and Gray 
1988). They discussed the structure of magnetic field based on their contemporaneous 
observations of high-precision broad-band linear polarimetry and circular 
spectropolarimetry in the 2017 season, 

Although these up-to-date polarimetric observations have deepened our understanding 
on the magnetic properties of $\xi$~Boo~A, we note that conventional variability 
studies of spectral line strengths/profiles of this star, such as conducted by 
Toner and Gray (1988) almost 3 decades ago, seem to have barely been done since then.
As individual lines have different sensitivities to changes in the physical 
condition (e.g., temperature), we may be able to get useful information by 
carefully examining the variations of diverse spectral lines together. 
Conveniently, a sufficient number of high-dispersion spectra of $\xi$~Boo~A
are at our disposal, which were obtained by observations covering $\sim 80$ nights 
in the period of 2008 winter through 2010 spring (including intensive observations
of $\sim 1$ week, where an iodine cell was also used for radial velocity analysis). 
Given this situation, we decided to make use of these data to study the nature of 
possible activity/inhomogeneity-related spectral variabilities, while examining 
the strengths/widths of many lines of different properties measured for each 
of the time-series spectra.

The points of interest upon which we wanted to check are as follows:
\begin{itemize}
\item
Do the atmospheric parameters spectroscopically determined based on Fe lines,
which should represent the mean values averaged over inhomogeneous stellar disk,
show notable variations?
\item
Do the indicators of chromospheric activity (such as the core emission of strong 
Ca~{\sc ii} line or He~{\sc i} 5876 line) show appreciable time variability?  
\item
What kind of results are obtained by precise radial velocity analysis?
Is any significant trend observed?
\item
Can we detect meaningful variations in spectral line strength/widths under 
inevitable influence of measurement noise? If so, are there any relations 
between these observables and relevant atmospheric parameters? 
\item
What about the influence of magnetic field on the spectrum? Is it possible to
find any dependence upon the effective Land\'{e} factor in the width/strength 
of spectral lines?
\item
Can we confirm the rotational modulation of spectrum variability corresponding
to the rotation period of 6.4~d reported by previous studies?
\end{itemize} 
 
The remainder of this article is organized as follows: 
We first explain our observational data in section~2 and 
spectroscopic determinations of atmospheric parameters in section~3.
Section~4 addresses the analysis of radial velocity variations
applied to the spectra obtained with I$_{2}$ cell. 
In section~5 are described our procedures of measuring broadening widths 
as well as equivalent widths of spectral lines, which are applied to 
selected 99 lines. The results of our analysis (especially in terms of 
the temperature sensitivity and the effect of magnetic field) are
discussed in section~6, where the detectability of rotational modulation 
is also mentioned based on our power spectrum analysis. 
The conclusions are summarized in section~7.

\section{Observational data}

\subsection{Observations with GAO/GAOES}

The main spectroscopic observations of $\xi$ Boo A were carried out on 49 nights 
from 2008 December through 2010 March with GAOES (Gunma Astronomical Observatory 
Echelle Spectrograph) installed at the Nasmyth Focus of the 1.5~m reflector 
of Gunma Astronomical Observatory. Setting the slit width at 1$''$, we could 
obtain spectra with a resolving power of $R \sim 70000$. In a night, we tried 
to make observations in two wavelength regions as much as possible,
4940--6810~\AA\ (33 orders, g-region) and 7550--9400~\AA (16 orders, i-region),
though the priority is on the former. 
As a result, while g-region spectra could be obtained for all the 49 nights, 
i-region spectra were eventually available only for 37 (out of 49) nights.
The journal of GAO/GAOES observations is summarized in table~1
and also illustrated in figure~1a.
Besides, we made special violet-blue region (3650--4830~\AA) observations 
on two nights of 2009 December 11 and 2010 January 6 (not included in table~1 
or figure~1a), which are only for the purpose of checking the core emission 
feature of Ca~{\sc ii} H+K lines.

\subsection{Observations with OAO/HIDES}

In addition, we also conducted intensive observations on 8 nights 
in 2010 April (27, 29, 30) and May (1, 2, 3, 4, 5) by using the 188~cm 
reflector along with HIDES (HIgh Dispersion Echelle Spectrograph) at 
Okayama Astrophysical Observatory. By using the slit of 200~$\mu$m
width, we could obtain spectra with a resolving power of $R \sim 70000$.
The resulting spectra turned out to cover the wavelength range of 
5000---8800~\AA\ (i.e., 5030--6260, 6300--7540, and 7600--8810~\AA\ by 
using three mosaicked CCDs).
The observations in a night were done in three time zones (a: early night, 
b: mid-night, and c: late night) separated with an interval of 3--4 hours.
Besides, two kinds of observations were done in each zone: (i) normal 
observation and (ii) observation while placing an iodine cell in front of
the entrance slit (in order to imprint I$_{2}$ molecular lines used for
precise radial velocity analysis). As a result of 22 observing opportunities
(1 + 3$\times$7; because on April 27 was observed only 1 time zone c),
44 (=22$\times$2) spectra were obtained in total. 
Table~2 gives the journal of these OAO/HIDES observations, which is
also graphically depicted in figure~1b.  

\subsection{Data reduction}

The reduction (bias subtraction, flat-fielding, aperture-determination, 
scattered-light subtraction, spectrum extraction, wavelength calibration, 
and continuum-normalization) of all the raw spectral data was performed 
by using the {\tt echelle} package of {\sc IRAF}.\footnote{IRAF is 
distributed by the National Optical Astronomy Observatories, 
which is operated by the Association of Universities for  Research  
in Astronomy, Inc., under cooperative agreement with the National 
Science Foundation.} As presented in tables 1 and 2, the mean S/N ratios of 
the resulting spectra widely differ from case to case ($\sim$~50--350)
but are typically $\sim$~100--200 on the average.
There is a tendency that S/N ratio in the near-IR region (i-region) 
is lower than the green--yellow region (g-region) because of
the lower sensitivity for the former (cf. table~1). Besides, S/N ratios of 
the spectra obtained with I$_{2}$ cell tend to be comparatively deteriorated 
due to the loss of light by the cell.

\subsection{Activity-sensitive lines}

As we are interested in examining whether lines sensitive to
chromospheric activity show any appreciable variability, we paid attention to 
the features of three lines (He~{\sc i} 5876, Ca~{\sc ii} 8542, and 
Ca~{\sc ii} 3934) in our spectra. 
Since the He~{\sc i} 5876 line is located in the spectral region contaminated by 
telluric water vapor lines, we removed these telluric lines (using the {\sc IRAF}
task {\tt telluric}) by dividing the spectra of relevant region 
by that of Regulus, which is a very rapid rotator 
($v_{\rm e}\sin i \simeq  300$~km~s$^{-1}$) and thus the spectral lines of stellar origin 
are almost smeared out. This process turned out reasonably satisfactory in most cases, 
though slight residuals occasionally remained unremoved, which appear as noises 
or continuum fluctuations on the order of a few per cent. 
The spectra in the neiborhood of these lines are overplotted in figure~2b,b$'$ 
(He~{\sc i} 5876), figure~2c,c$'$ (Ca~{\sc ii} 8542), and figure~2d (Ca~{\sc ii} 3934),
where the reference solar spectra are also depicted for comparison. 
We can see from these figures that, although these lines show signs of conspicuously 
high  chromospheric activity of $\xi$~Boo~A compared to the Sun (i.e., stronger 
He~{\sc i} 5876 absorption and higher Ca~{\sc ii} 8542/3934 core emission), 
appreciable time variations are not recognized. 

\section{Spectroscopic determination of atmospheric parameters}

\subsection{Absolute parameter determination}

The atmospheric parameters [$T_{\rm eff}$ (effective temperature), $\log g$ 
(logarithmic surface gravity), $v_{\rm t}$ (microturbulence), and [Fe/H] 
(logarithmic Fe abundance relative to the Sun; often denoted as ``metallicity'') 
were spectroscopically determined by using Fe~{\sc i} and Fe~{\sc ii} lines 
in the same way as done by Takeda et al. (2005; see subsection~3.1 therein). 
The equivalent widths of these Fe lines were measured in the conventional manner 
by applying the Gaussian-fitting method.\footnote{Specifying the relevant 
wavelength range [$\lambda_{1}$, $\lambda_{2}$] and the continuum level ($f_{\rm cont}$), 
we determined such a Gaussian function that best fits the line profile ($f_{\lambda}$) 
by applying the non-linear least-squares fitting algorithm where three parameters 
(peak value, $e$-foldig width, and wavelength shift) were adjusted.}
In order to achieve consistency between the results obtained by spectra of 
different observatories, only lines in the 5030--6800~\AA\ region (i.e., 
common to the OAO/HIDES spectra and g-region of GAO/GAOES spectra) were employed, 
and we restricted to using only lines not stronger than 100~m\AA\ as in 
Takeda et al. (2005).
In the online material are presented the measured equivalent widths along 
with the corresponding Fe abundances (tableE1.dat) and the resulting parameters
($T_{\rm eff}$, $\log g$, $v_{\rm t}$, and [Fe/H]) (tableE2.dat) 
for each of the 71 (= 49 GAO + 22 OAO) spectra.  The typical statistical errors
(see subsection~5.2 of Takeda et al. 2002) involved with these solutions 
are $\sim 10$~K, $\sim$~0.03~dex, $\sim 0.1$~km~s$^{-1}$, and $\sim 0.02$~dex,
respectively.
The distribution histograms of these absolute parameters are presented in figure~3.
Given that these results show more or less spreads, it is convenient to assign 
``standard'' parameters\footnote{We defined these representative parameters 
rather arbitrarily, based on which the standard model atmosphere used for computing 
$F^{0}(\lambda)$ was constructed (cf. subsection~5.2). They do not have any other 
special meaning (such as the mean parameters averaged over the phases).} 
of $\xi$~Boo~A for reference, for which we take 
 (5527~K, 4.60, 1.10~km~s$^{-1}$, and $-0.13$~dex) derived by Takeda et al. (2005),
as indicated by downward arrows in figure~3.
       
\subsection{Differential parameter determination}

Next, we further established the differential parameters 
($\Delta T_{\rm eff}$, $\Delta \log g$, $\Delta v_{\rm t}$, and $\Delta$[Fe/H]) 
relative to the fiducial values (for which we took those corresponding to
the first observation for each observatory; i.e., 20081210g for GAO/GAOES
and 0427c for OAO/HIDES) by applying the method of purely differential analysis
developed by Takeda (2005), as given in tableE2.dat.
While such determined differential parameters should be comparatively more precise 
than the absolute parameters, we can see that they ($\Delta p$) are reasonably 
correlated with the simple difference of absolute parameters ($p - p^{0}$; 
where fiducial values are denoted by superscript `0'), as demonstrated in figure~4.

\section{Precision analysis of radial velocity variation}

Since the spectra observed with I$_{2}$ cell were obtained along with 
the normal spectra in each of the time zones for the case of OAO/HIDES 
observations, we made use of them to precisely evaluate the radial 
velocity variations relative to the fiducial first I$_{2}$ spectrum (0427i2c)
while following the procedure described in section~4 of Takeda et al. (2002)
(cf. equations (10)--(12) therein), where we employed ``0427c'' as 
the template ``pure star'' spectrum. Only the spectral range of 5030--6230~\AA\
was used, where I$_{2}$ molecular lines are appreciably observed in absorption. 
The results of analysis are summarized in tableE3.dat of the online material.
The differential heliocentric radial velocities relative to 0427i2c 
(typical probable errors being several to $\ltsim 10$~m~s$^{-1}$) are plotted
against the observed time in figure~5 (panel a), where the runs of 
$\Delta T_{\rm eff}$ (panel b) and mean S/N ratio (panel c) are also shown.

It can be seen from figure~5a that the dispersion of radial velocity variations is 
$\sim \pm 50$~m~s$^{-1}$ with a broad dip around HJD~2455321--2455322 (0503i2a--0504i2c).
Regarding the outlier point at HJD 2455319.3 (0502i2c), we do not have 
much confidence about its credibility, because the relevant spectrum is 
of the lowest S/N ratio ($\sim 60$) and the probable error is accordingly
larger (15~m~s$^{-1}$). As such, it may be premature to mention the existence 
of any correlation between the radial velocity and $\Delta T_{\rm eff}$ 
from these figures alone.

\section{Measurement of spectral lines}

\subsection{Target lines}

As to the lines to be analyzed, we adopted 99 lines (satisfying 
the criterion of not seriously blended with other stellar lines or 
telluric lines) selected from two wavelength regions. 

The first is the orange region of 6000--6260~\AA, which is known to contain 
a number of high-quality lines used for line profile studies (e.g., 
Toner \& Gray 1988). By consulting Takeda and UeNo's (2019) line list,
we selected 55 lines from this region, which are mostly due to neutral 
species (Na~{\sc i}, Si~{\sc i}, Ca~{\sc i}, Ti~{\sc i}, V~{\sc i}, 
Fe~{\sc i}, Ni~{\sc i}) though several are of ionized species
(Sc~{\sc ii}, Fe~{\sc ii}).

The second is the near-IR region (7560--8920~\AA). This is because 
we wanted to study the Zeeman effect of magnetic field, for which 
using lines of longer wavelength is more advantageous. 
By comparing the theoretical synthesized spectrum (computed by using 
the standard model atmosphere defined in subsection~3.1 along with
the atomic line data compiled by Kurucz \& Bell 1995) with the observed 
spectrum, suitable lines were sorted out, which we intentionally 
restricted to Fe~{\sc i} lines. The only exception was the O~{\sc i} 7774
line (middle line of the triplet), which was specially included.
As a result, 44 lines were chosen from this near-IR region.

The list of these 99 lines is given in table~3, where the 
relevant atomic data are also presented.  
Note that near-IR lines at 7568--7586~\AA\ or 8824--8920~\AA\
could not be measured for the case of OAO/HIDES spectra 
because they fall outside of the CCD format. 

\subsection{Method of analysis}

Regarding the measurement of physical quantities (equivalent widths, 
line broadening widths) of these spectral lines, our adopted method is based 
on the fitting of observed spectrum with a theoretically modeled profile, 
as recently done by Takeda and UeNo (2019) for their intensity spectrum 
analysis on the solar disk.
The modeling of line-profile is almost the same as described in section~3 of 
that paper, except that flux (i.e., angle-averaged specific intensities)
is involved in this case.
The observed flux profile $F(v)$\footnote{In equations (1) and (3), 
the profile point is specified by $v$ (velocity variable) for simplicity 
instead of $\lambda$ (wavelength).} is expressed as
\begin{equation}
F(v) = F^{0}(v) \otimes K(v),
\end{equation}
where `$\otimes$' means ``convolution.''
Here, $F^{0}(v)$ is the unbroadend emergent flux profile at the surface, 
which is written by the formal solution of radiative transfer as 
\begin{equation}
F^{0}(\lambda) \propto \int_{0}^{\infty} S_{\lambda}(t_{\lambda}) E_{2}(t_{\lambda}) 
{\rm d}t_{\lambda},
\end{equation}
where $S_{\lambda}$ is the source function, $t_{\lambda}$ is the optical depth
in the vertical direction, and $E_{2}$ is the exponential integral function of
2nd order (see, e.g., Gray 2005). Further, $K(v)$ is the Gaussian macrobroadening function
with $e$-folding half-width of $v_{\rm M}$\footnote{
Note that $v_{\rm M}$ is expressed by the root-sum-square of three broadening widths:
(i) instrumental broadening ($v_{\rm ip} \simeq (c/R)/(2\sqrt{\ln 2}$); 
$c$ is the velocity of light, and $R$ is the spectral resolving power),
(ii) rotational broadening ($v_{\rm rt}$), and 
(iii) macroturbulence broadening ($v_{\rm mt}$) as
$v_{\rm M}^{2} = v_{\rm ip}^{2} + v_{\rm rt}^{2} + v_{\rm mt}^{2}$. 
See sub-subsection 4.2.1 and footnote~12 of Takeda, Sato, and Murata (2008) or 
Appendix 3 of Takeda and UeNo (2017) for more details.}
\begin{equation}
K(v) \propto \exp[-(v/v_{\rm M})^2].
\end{equation} 

As for the calculation of $F^{0}(\lambda)$, we adopted Kurucz's 
(1993) ATLAS9 model atmosphere corresponding to the standard parameters
of $\xi$~Boo~A ($T_{\rm eff}$ = 5527~K, $\log g = 4.60$, [Fe/H] = $-0.13$) 
with a microturbulent velocity of $\xi$ = 1.1~km~s$^{-1}$ (cf. subsection~3.1)
while assuming LTE. We adopted the algorithm described in Takeda (1995) to 
search for the best-fit theoretical profile, while varying three parameters 
[$\log\epsilon$ (elemental abundance), $v_{\rm M}$ (macrobroadenig
width), and  $\Delta\lambda_{\rm r}$ (wavelength shift)] 
for this purpose. As to the atomic parameters of each spectral line 
($gf$ values, damping constants), we exclusively adopted the values 
presented in Kurucz and Bell's (1995) compilation. 
The background opacities were included as fixed by assuming 
the scaled solar abundances according to the metallicity.

After the solution has been converged, we can use the resulting abundance
solution ($\log\epsilon$) to compute the corresponding equivalent 
width ($W$)  with the help of Kurucz's (1993) WIDTH9 program:
\begin{equation}
W \equiv \int R^{0}(\lambda) {\rm d}\lambda
\end{equation}
where 
$R^{0}(\lambda)$ is the line depth of theoretical 
flux profile with respect to the continuum level 
[$R^{0}(\lambda) \equiv 1 - F^{0}(\lambda)/F^{0}_{\rm cont}$]
and integration is done over the line profile.

\subsection{Results}

In order to check the reliability of this procedure, we measured the equivalent 
widths of these 99 lines on the representative GAOES spectra (20081210g, 20081219i) 
by using the conventional Gaussian-fitting method and compared them with those 
derived by the approach of modeled-profile fitting adopted in this study. 
This comparison is illustrated in figure~6, where we can confirm a reasonable consistency.

The resulting values of $W$ and $v_{\rm M}$ measured for each line 
on each spectrum are presented in tableE4.dat of the online material.
In figure~7 are plotted the runs of $\log W$ and $\log v_{\rm M}$
against the observed dates for selected 5 representative Fe~{\sc i} lines of
different strengths ($W \sim$~12--171~m\AA). 
The mean values ($\langle \log W \rangle$, $\langle v_{\rm M} \rangle$)
averaged over each of the spectra and their standard deviations are
given in table~3 for all of the 99 lines.

\section{Discussions} 

\subsection{Detectability of line-strength variation}

Having taken a glance at the results obtained in section 5, we realized that 
spectral variabilities are so small that they are not necessarily easy to detect. 
Especially, since variability signals in equivalent widths can be obscured 
by random noises depending upon the line strengths, it is worthwhile to 
examine the detectability of $W$ variation under the influence 
of measurement errors. 

The uncertainties in the equivalent width ($\delta W$) 
due to random noises can be estimated 
by invoking the relation derived by Cayrel (1988),
\begin{equation}
\delta W \simeq 1.6 (w \delta x)^{1/2} \epsilon,
\end{equation}
where $\delta x$ is the pixel size ($\simeq 0.03$~\AA\ for the case of
our observation), $w$ is the width of a line (for which we may roughly set 
$\sim 0.2$~\AA\ for $\xi$~Boo~A), and $\epsilon \equiv ({\rm S/N})^{-1}$.
Accordingly, the error in $\log W$ can be written as
\begin{equation}
\delta \log W =  \frac{\ln (1+\delta W/W)}{\ln 10} 
\simeq \frac{\delta W/W}{\ln 10},
\end{equation} 
which indicates that $\delta \log W$ progressively
increases with a decrease in $W$. 

The $\sigma_{\log W}$ values (standard deviations) computed for each of the 
99 lines are plotted against the corresponding $\langle \log W \rangle$
(mean equivalent widths) in figure~8a, where the predicted relations based on 
equations (5) and (6) are also shown by dashed lines for four S/N values 
(50, 100, 200, and 400).
This figure suggests that the observed behavior of $\sigma_{\log W}$ 
is similar to the expected trend due to random noises especially at 
$\log W \ltsim 1.5$, which means that extracting useful information 
is difficult from such weak lines. However, considering that the typical S/N 
ratios of our spectra are $\sim$~100--200, we recognize that a significant fraction 
of the $\sigma_{\log W}$ values are above the noise-limited relation
for lines of medium-to-large strengths ($\log W \gtsim 1.5$).
Accordingly, we may regard that the dispersion of $\log W$ shown by such 
stronger lines are real, which should have stemmed from actual stellar variability. 

\subsection{Temperature sensitivity}

Now that fluctuations of $\log W$ exhibited by moderately-strong and strong lines 
are considered to be real, the next task is to trace down their physical cause. 
We may expect that the important parameter affecting the line strengths would be $T$ 
(temperature), as the diversified trends of solar center--limb variation in $W$  
are essentially determined by the sensitivity to $T$ differing from line to line 
(cf. Takeda \& UeNo 2019). 
As done by Takeda and UeNo (2019), we evaluated the $T$-sensitivity indicator 
$K (\equiv {\rm d}\log W/{\rm d}\log T)$ as follows:
\begin{equation}
K \equiv \frac{(W^{+100} - W^{-100})/W}{(+100-(-100))/5527},
\end{equation}
where $W^{+100}$ and $W^{-100}$ are the equivalent widths computed (with the 
same $\log\epsilon$ solution reproducing the observed equivalent width $W$) 
by two model atmospheres with only $T_{\rm eff}$ being perturbed by $+100$~K 
($T_{\rm eff} = 5427$~K) and  $-100$~K ($T_{\rm eff} = 5627$~K), respectively
(while other parameters are kept the same as the standard values; cf. subsection~3.1).
The resulting $K$ for each line is presented in table~3.

The behavior of $K$ is distinctly different depending on whether the 
considered species is of minor population ($K < 0$) or major population 
($K > 0$) as briefly summarized below (see Takeda \& UeNo 2019 for more details):
\begin{itemize}
\item
In the weak-line case, $K$ follows the analytical relations of 
$K^{\rm minor} \simeq -11604 (\chi^{\rm ion} -\chi_{\rm low})/T$ (minor population)
and  
$K^{\rm major} \simeq +11604 \chi_{\rm low}/T$ (major population),
where $\chi^{\rm ion}$ (ionization potential) and $\chi_{\rm low}$ (lower excitation
potential) are in unit of eV and $T$ is in K.
\item
As lines get stronger and more saturated, $K$ tends to become progressively 
smaller than that given by these analytical relations; i.e., stronger lines 
tend to show smaller $T$-sensitivity compared to weaker lines at the same 
potential energy.
\end{itemize}
The $K$ values of 99 lines are plotted against $\chi^{\rm ion} -\chi_{\rm low}$
(minor population species) or $\chi_{\rm low}$ (major population species) 
in figure 9, where we can confirm these characteristics.

Accordingly, for example, if we are to find lines of stronger $T$-sensitivity 
(i.e., larger $|K|$), (i) weaker lines of (ii) larger $\chi^{\rm ion} -\chi_{\rm low}$ 
(minor population) or $\chi_{\rm low}$ (major population) should be preferred. 
However, since weak lines severely suffer from the effect of random 
noises and unsuitable as shown in subsection~6.1, choosing line of
moderate strengths (e.g., several tens of m\AA) would be a practically 
reasonable choice for examining the effect of $T$. 

We selected 8 lines of $W \sim $~25--50~m\AA\ with various $K$ values ranging 
from $-11.8$ to $+6.6$, and plotted their $\log W$ values derived from each 
of the available spectra against the corresponding $\log T_{\rm eff}$ in figure~10. 
We can see from this figure that the observed $\log W$ vs. $\log T_{\rm eff}$ 
trends of these lines of different $T$-sensitivity are almost consistent with 
those expected from the $K$ values (i.e., positive/negative gradient for 
positive/negative $K$). This fact implies that the spectrum variability 
is mainly determined by the mean surface temperature averaged over the
stellar disk showing structural inhomogeneities.

\subsection{Line-broadening width}

We then turn our attention to $v_{\rm M}$ (macrobroadening width),
in which various broadening components are involved, such as
instrumental broadening ($v_{\rm ip}$), macroturbulence ($v_{\rm mt}$)
and rotational broadening ($v_{\rm rt}$) as mentioned in footnote~5.
Figure~8c indicates that $v_{\rm M}$ does not exhibit any clear dependence 
upon line strengths but distributes in the range of 
$0.6 \ltsim \log v_{\rm M} \ltsim 0.7$. Is this dispersion real? 

Here, it is meaningful to check whether any effect of magnetic field 
can be observed in this parameter, because the effect of Zeeman broadening 
can also be formally included such as like ``line-dependent 
macroturbulence.'' Interestingly, we recognize in figure~8d some sign of 
positive correlation between $\log v_{\rm M}$ and $g_{\rm eff}^{\rm L}$ 
(effective Land\'{e} factor), which appears comparatively more manifest
for lines of medium--large strengths. 
Moreover, figure~8b reveals that a similar correlation is observed in 
$\sigma_{\log W}$ of stronger lines; i.e., standard deviation of $\log W$ 
tends to progressively increase with $g_{\rm eff}^{\rm L}$
for lines of $W \gtsim 100$~m\AA, indicating that such saturated lines
suffer magnetic intensification.
According to these consequences, we can state that the widths and strengths 
of spectral lines in $\xi$~Boo~A are possibly affected by the existence of 
magnetic field (to line-by-line different extents depending the individual 
Zeeman sensitivity and line strengths).
 
In this respect, it is worth to mention that Morgenthaler et al. (2012)
reported a significant long-term variation in the width of the Zeeman-sensitive
($g_{\rm eff}^{\rm L} = 2.5$) Fe~{\sc i} line at 8468.4~\AA,
which is fairly well correlated with the change of Ca~{\sc ii} index.
Their figure~7 (lower panel) suggests (though the scatter is large) that the 
width of this line showed a decreasing tendency from late 2008 (through mid-2009) 
to early 2010, but then turned to exhibit an increasing trend toward mid-2010. 
An inspection of our figure~7e indicates that an appreciable change 
in support of their observation can be detected in the width of this line 
(e.g., systematic decrease from the end of 2009 (around HJD~2455200) to 
early 2010 (around HJD~2455300). Unfortunately, since the near-IR data are 
lacking in the perid of HJD~2455000--2455200, it is hardly possible to discuss 
its variation in detail based on our data alone. In any event, such a long-term 
variability in the width of this Fe~{\sc i} line implies that magnetic areas
exist on the surface of $\xi$~Boo~A and their condition undergoes changes 
with time due to variations of stellar activity.
 
Finally, $v_{\rm e}\sin i$ (projected equatorial rotational velocity) 
can be estimated from $v_{\rm M}$. Since we see  
$\log v_{\rm M} \simeq 0.6$ at $g_{\rm eff}^{\rm L} \rightarrow 0$ (cf. figure~8d), 
we may regard $v_{\rm M} \simeq 4.0$~km~s$^{-1}$ at the non-magnetic case.
Let us recall that $v_{\rm M}$ is expressed as root-sum-square of
$v_{\rm ip}$ (instrumental broadening), $v_{\rm mt}$ (macroturbulence
broadening), and $v_{\rm rt}$ (rotational broadening), as described in footnote~5.
Considering that 
$v_{\rm ip} \simeq (3\times 10^{5}/70000)/(2\sqrt{\ln 2}) \simeq 2.6$~km~s$^{-1}$
 and $v_{\rm mt}$ = 1.5~km~s$^{-1}$ for $T_{\rm eff} \simeq 5500$~K by using 
equation (A3) of Takeda and UeNo (2017), we have 
$v_{\rm rt}$ = $\sqrt{4.0^{2}-2.6^{2}-1.5^{2}}$ = 2.6~km~s$^{-1}$. 
Adopting the conversion relation of $v_{\rm e}\sin i = v_{\rm rt}/0.94$
(cf. footnote~12 in Takeda et al. 2008), we finally obtain
$v_{\rm e}\sin i$ = 2.8~km~s$^{-1}$ for $\xi$~Boo~A.\footnote{As seen from 
the large scatter of $\langle \log v_{\rm M}\rangle$ vs. $g_{\rm eff}^{\rm L}$ 
relation (figure~8d), we assume that the adopted $\langle \log v_{\rm M}\rangle = 0.6$~dex 
(in the limit of $g_{\rm eff}^{\rm L} \rightarrow 0$) may involve an uncertainty 
on the order of $\sim \pm 0.05$~dex. This corresponds to an error of 
$\sim \pm$~0.8--0.9~km~s$^{-1}$ in the resulting $v_{\rm e}\sin i$ of 2.8~km~s$^{-1}$.} 
This value is in remarkable agreement with 2.9~km~s$^{-1}$ derived by 
Gray (1984), while somewhat smaller than $4.6 (\pm 0.4)$~km~s$^{-1}$
compiled by Marsden et al. (2014).

\subsection{Comparison with Toner and Gray (1988)}

As mentioned in section~1, it was one of our motivations to confirm the results 
of Toner and Gray (1988), who concluded by careful analysis of spectral line profiles
(1) the existence of a large patch and (2) the rotational modulation 
period of 6.43~d. We realized, however, that it is hardly possible to study
asymmetries of line profiles (i.e., bisector analysis) such as done by 
Toner and Gray (1988), presumably because our observational data ($R \sim 70000$; 
typical S/N $\sim$~100--200) are of considerably lower quality compared to theirs
($R \sim 86000$; typical S/N $\sim$~300--600; cf. their table~1).
Yet, their result that line profile asymmetry varies in the velocity span of 
$\sim 100$~m~s$^{-1}$ (cf. their figure~6) may be regarded as consistent with the 
variability range of differential radial velocities ($\sim 100$~m~s$^{-1}$; cf. figure~5a).

Since our line-strength measurements should be comparatively more reliable, we here
focus on the equivalent width ratio of Fe~{\sc ii} 6247.56 and V~{\sc i} 6251.83  
lines, for which Toner and Gray (1988) detected cyclic variations ($\sim 10$\%) 
with a 6.4~d period (cf. their figure~3) similar to the behavior of profile asymmetry. 
This line pair is an ideal indicator of $T_{\rm eff}$ variation, 
because these two lines show large $T$-sensitivities of opposite sense 
(i.e., $K = +3.41$ for Fe~{\sc ii} 6247.56 and $K = -11.79$ for V~{\sc i} 6251.83).
Actually, $W$(V~{\sc i}) and $W$(Fe~{\sc ii}) are anticorrelated (figure~11a),
and the $W$(V~{\sc i})/$W$(Fe~{\sc ii}) ratio shows a decreasing tendency 
with an increase in $T_{\rm eff}$ (figure~11c). We carried out a power spectrum 
analysis on the dataset of $W$(V~{\sc i})/$W$(Fe~{\sc ii}) during the 
period of HJD~2454810--2455010\footnote{The reason why we restricted ourselves 
to the data in this time span is that 
the sample data should be not only wealthy but also uniform and of wide-coverage
(with respect to time) in order to make power spectrum analysis successful.}
(end-2008 through mid-2009; see the range indicated by dashed line in figure~11b 
or figure~11d) to see whether any cyclic pattern is contained therein. 
The Fortran subroutine {\tt PERIOD.FOR} (for computing power spectrum of unevenly 
sampled data; cf. Press et al. 1992) was used for this purpose.

The resulting power spectrum for $W$(V~{\sc i})/$W$(Fe~{\sc ii}) is depicted 
by blue line in figure~11e, where those calculated for $T_{\rm eff}$
(subsection~3.1) and $\Delta T_{\rm eff}$ (subsection~3.2) are also shown  
in red and pink lines, respectively.
As clearly seen from this figure, we can not detect any such strong single 
peak corresponding to 6.43~d period as reported by Toner and Gray (1988; cf. 
their figure~2) based on their 1986 observations. What we can barely recognize 
in figure~11e is several rather weak peaks, among which notable ones 
commonly observable for $W$(V~{\sc i})/$W$(Fe~{\sc ii}), $T_{\rm eff}$, 
and $\Delta T_{\rm eff}$ are those at the periods of $\sim 6.1$~d 
and $\sim 7.4$~d.

The natural interpretation for the cause of this discrepancy would be that 
the surface features of $\xi$~Boo~A at the time of their observations
in 1986 were considerably different from our case (observed mostly in 
early-to-mid 2009). The lack of strong peak in our data may suggest 
that star spots (or faculae) are widely disassembled over the stellar surface, 
rather than a large star patch located at high latitude concluded by Toner and Gray (1988).
The fact that weak multi-peaks are observed in figure~11e may also be explained
by this picture, since rotation of $\xi$~Boo~A is expected to be strongly 
differential according to Morgenthaler et al. (2012).
In short, activity-related inhomogeneities on the surface of $\xi$~Boo~A 
would have been rather complex and dispersed at the time of our observations.

\section{Summary and conclusion}

As the solar-type star $\xi$~Boo~A is known to be chromospherically active, 
it is expected to show appreciable spectral variations caused by rotation-modulated 
surface inhomogeneities or long-term changes of magnetic activity.
However, ever since Toner and Gray (1988) detected significant variabilities in 
line strengths as well as profile asymmetries and concluded the existence of
a large star patch on the surface of this star, similar studies targeting 
individual spectral lines have scarcely been published over these 30 years,
despite that several investigators have made progress in clarifying the nature 
of magnetic fields based on modern polarimetric observations.
 
Motivated by this situation, we decided to carry out a spectroscopic study of $\xi$~Boo~A
based on the available 71 time-series high-dispersion spectra obtained at Gunma 
Astronomical Observatory and Okayama Astrophysical Observatory in 2008 December 
though 2010 May. Our aims were (1) to detect any variability in spectroscopically 
determined parameters as well as measurable quantities (strengths or widths) 
of many spectral lines, and (2) to understand the physical cause of such variation. 

We first checked the features of spectral lines (He~{\sc i} 5876 absorption, 
core emission of Ca~{\sc ii}~8542/3934) which are known to be especially 
sensitive to chromospheric activity. These indicators were confirmed to be
manifestly strong as compared to the Sun (indicating higher activity).
However, we could not recognize any clear time variability in them. 

The equivalent widths of many Fe lines in the 5030--6800~\AA\ region were measured
for each spectrum by applying the conventional Gaussian fitting method. 
Based on these data, the atmospheric parameters ($T_{\rm eff}$, $\log g$, $v_{\rm t}$, 
and [Fe/H]) were spectroscopically determined as done by Takeda et al. (2005) 
(absolute values) and Takeda (2005) (differential values), which revealed that 
$T_{\rm eff}$ (mean temperature averaged over the disk) fluctuates by $\sim \pm$~30--40~K. 

Since spectra with imprinted I$_{2}$ molecular lines (usable for precise wavelength
calibration) were also obtained in the intensive observations in 2010 late April 
and early May at OAO, we made use of them to examine radial velocity variations 
over this time span of $\sim 1$ week. The resulting differential radial velocities 
show a dispersion of $\sim \pm 50$~m~s$^{-1}$ around the mean with some systematic 
trend (i.e., a broad dip), though they do not appear to definitely correlate with 
the change of $T_{\rm eff}$.

By applying the efficient method (Takeda \& UeNo 2019), which fits the observed line 
profile with the parameterized model profile, line-broadening widths ($v_{\rm M}$) 
and the equivalent widths ($W$) of 99 high-quality lines selected from the 
orange region (6000--6260~\AA) and near-IR region (7560--8920~\AA) were measured.

Regarding equivalent widths, while the changes are difficult to distinguish from 
random noises for weak lines, we could detect meaningful fluctuations for 
moderate-strength lines. Plotting the $\log W$ values of representative lines 
(with $W$ of several tens m\AA) showing different $T$-sensitivities 
against $\log T_{\rm eff}$, we could confirm that $\log W$ and $\log T_{\rm eff}$
well correlate with each other in just the expected manner (i.e., positive/negative 
correlation for major/minor population species). This means that a change in the mean 
temperature averaged over the disk (of inhomogeneous surface structure) is mainly
responsible for the line-strength variation.  

As to $v_{\rm M}$, this line-broadening width parameter was found to show an increasing
trend with $g_{\rm eff}^{\rm L}$ (effective Land\'{e} factor) for lines of moderate or
large strengths. We also note that $\sigma_{\log W}$ (standard deviation of $\log W$) 
also tends to grow with $g_{\rm eff}^{\rm L}$ for strong saturated lines, which implies
the existence of magnetic intensification. These observational facts suggest that the 
spectrum (and its variability) of $\xi$~Boo~A is appreciably influenced by magnetic 
fields, depending on the Zeeman sensitivity of each line.

For the purpose of confirming Toner and Gray's (1988) conclusion, we paid our
attention to the equivalent width ratio of Fe~{\sc ii} 6247.56 and V~{\sc i} 6251.83 
lines as they did.  However, our power spectrum analysis applied to this 
$W$(V~{\sc i})/$W$(Fe~{\sc ii}) ratio (as well as to $T_{\rm eff}$ and $\Delta T_{\rm eff}$) 
could not detect any strong peak at the period of 6.4~d such as that found by 
Toner and Gray (1988) based on their 1986 observations, though several peaks were 
weakly recognized (e.g., at $\sim 6.1$~d and $\sim 7.4$~d).
This may imply that surface inhomogeneities of $\xi$~Boo~A at the time of our 
observations (mainly in 2009) were not so much simple (like a large star patch) 
as rather complex (e.g., intricate aggregate of spots and faculae).  

\bigskip


\newpage
\onecolumn

\setcounter{table}{0}
\begin{table*}
\begin{minipage}{180mm}
\scriptsize
\caption{Spectra of GAO/GAOES observations from 2008 December through 2010 March.}
\begin{center}
\begin{tabular}{ccccccc}\hline
code & HJD & $\langle$S/N$\rangle$ & &  code & HJD & $\langle$S/N$\rangle$ \\
(1) & (2) & (3) &  & (4) & (5) & (6) \\
\hline
\multicolumn{3}{c}{[5000--6800 \AA\ region]} & & \multicolumn{3}{c}{[7500--9300 \AA\ region]}\\
20081210g & 4811.306 & 189 &  &      ---       &    ---       &  ---    \\
20081219g & 4820.273 & 174 &  &   20081219i &  4820.326 &  139 \\
20081220g & 4821.267 & 161 &  &   20081220i &  4821.317 &  135 \\
20090106g & 4838.233 & 194 &  &   20090106i &  4838.283 &  133 \\
20090107g & 4839.210 & 190 &  &   20090107i &  4839.245 &  109 \\
20090125g & 4857.254 & 225 &  &   20090125i &  4857.302 &  131 \\
20090126g & 4858.226 & 188 &  &      ---       &    ---       &  ---    \\
20090202g & 4865.190 & 218 &  &   20090202i &  4865.239 &   84 \\
20090206g & 4869.186 & 227 &  &   20090206i &  4869.220 &  101 \\
20090208g & 4871.241 & 258 &  &   20090208i &  4871.282 &  149 \\
20090210g & 4873.264 & 227 &  &      ---       &    ---       &  ---    \\
20090212g & 4875.267 & 100 &  &      ---       &    ---       &  ---    \\
20090214g & 4877.230 & 281 &  &   20090214i &  4877.271 &  131 \\
20090217g & 4880.192 & 222 &  &   20090217i &  4880.240 &  112 \\
20090221g & 4884.277 & 259 &  &   20090221i &  4884.327 &  104 \\
20090301g & 4892.255 & 148 &  &   20090301i &  4892.290 &   44 \\
20090302g & 4893.136 & 131 &  &   20090302i &  4893.169 &   33 \\
20090310g & 4901.219 & 225 &  &   20090310i &  4901.285 &  160 \\
20090320g & 4911.188 & 142 &  &   20090320i &  4911.230 &   76 \\
20090321g & 4912.167 & 186 &  &      ---       &    ---       &  ---    \\
20090409g & 4931.108 & 274 &  &   20090409i &  4931.061 &  176 \\
20090410g & 4932.148 & 349 &  &   20090410i &  4932.182 &  235 \\
20090411g & 4933.153 & 147 &  &   20090411i &  4933.178 &  122 \\
20090415g & 4937.126 & 221 &  &   20090415i &  4937.171 &  147 \\
20090419g & 4941.156 & 167 &  &   20090419i &  4941.196 &  121 \\
20090422g & 4944.163 & 165 &  &      ---       &    ---       &  ---    \\
20090426g & 4948.217 & 207 &  &   20090426i &  4948.254 &  113 \\
20090427g & 4948.991 & 153 &  &   20090427i &  4949.044 &  138 \\
20090508g & 4960.089 & 176 &  &   20090508i &  4960.132 &  131 \\
20090513g & 4965.005 & 172 &  &   20090513i &  4965.058 &  119 \\
20090519g & 4970.990 & 196 &  &   20090519i &  4971.043 &  147 \\
20090525g & 4976.987 & 117 &  &   20090525i &  4977.039 &  131 \\
20090526g & 4977.983 &  63 &  &      ---       &    ---       &  ---    \\
20090601g & 4984.174 & 230 &  &   20090601i &  4984.127 &  144 \\
20090625g & 5007.973 &  81 &  &      ---       &    ---       &  ---    \\
20090626g & 5009.109 & 146 &  &   20090626i &  5009.139 &  123 \\
20090907g & 5081.929 & 150 &  &      ---       &    ---       &  ---    \\
20090916g & 5090.978 &  56 &  &      ---       &    ---       &  ---    \\
20090917g & 5091.923 & 117 &  &      ---       &    ---       &  ---    \\
20091219g & 5185.307 & 204 &  &      ---       &    ---       &  ---    \\
20100116g & 5213.267 & 173 &  &   20100116i &  5213.313 &  117 \\
20100124g & 5221.251 & 155 &  &   20100124i &  5221.283 &  113 \\
20100126g & 5223.280 & 199 &  &   20100126i &  5223.325 &  133 \\
20100203g & 5231.143 & 120 &  &   20100203i &  5231.188 &  118 \\
20100204g & 5232.220 & 200 &  &   20100204i &  5232.256 &  129 \\
20100219g & 5247.226 & 164 &  &   20100219i &  5247.258 &  118 \\
20100222g & 5250.132 & 139 &  &   20100222i &  5250.170 &  139 \\
20100313g & 5269.266 & 127 &  &   20100313i &  5269.309 &   89 \\
20100319g & 5275.169 &  96 &  &   20100319i &  5275.214 &  103 \\
\hline
\end{tabular}
\end{center}
Columns (1) and (4) --- Spectrum code indicating the observed date and the
wavelength region. For example, ``20081210g'' is the data covering the
g-region (5000--6800 \AA) observed on 2008 December 10 (UT),
while ``20100319i'' is the data covering the i-region (7500--9300 \AA)
observed on 2010 March 19 (UT).
Columns (2) and (5) --- Heliocentric Julian day ($- 2450000$) corresponding 
to the observed time.
Columns (3) and (6) --- Mean signal-to-noise ratio computed as 
$\sqrt{\langle c \rangle}$, where $\langle c \rangle$ is the mean photoelectron 
counts of the whole echelle data (comprising 33 and 16 orders for g- and i-region, 
respectively) evaluated by the {\tt imstatistics} task of {\sc IRAF}.
\end{minipage}
\end{table*}

\setcounter{table}{1}
\begin{table*}
\begin{minipage}{180mm}
\scriptsize
\caption{Spectra of OAO/HIDES observations in 2010 late April and early May.}
\begin{center}
\begin{tabular}{ccccccc}\hline
code & HJD & $\langle$S/N$\rangle$ & &  code & HJD & $\langle$S/N$\rangle$ \\
(1) & (2) & (3) &  & (4) & (5) & (6) \\
\hline
\multicolumn{3}{c}{[normal observation]} & & \multicolumn{3}{c}{[observation with I$_{2}$ cell]}\\
0427c &  5314.281 &  238 &  &   0427i2c & 5314.301 &  180 \\
0429a &  5315.987 &  289 &  &   0429i2a & 5316.005 &  192 \\
0429b &  5316.172 &  197 &  &   0429i2b & 5316.191 &  194 \\
0429c &  5316.312 &  195 &  &   0429i2c & 5316.325 &  172 \\
0430a &  5316.975 &  148 &  &   0430i2a & 5317.006 &  130 \\
0430b &  5317.131 &  283 &  &   0430i2b & 5317.144 &  184 \\
0430c &  5317.309 &  271 &  &   0430i2c & 5317.324 &  160 \\
0501a &  5317.989 &  216 &  &   0501i2a & 5318.006 &  153 \\
0501b &  5318.131 &  229 &  &   0501i2b & 5318.143 &  148 \\
0501c &  5318.310 &  279 &  &   0501i2c & 5318.325 &  194 \\
0502a &  5318.994 &  197 &  &   0502i2a & 5319.012 &  125 \\
0502b &  5319.138 &  226 &  &   0502i2b & 5319.154 &  107 \\
0502c &  5319.311 &   76 &  &   0502i2c & 5319.327 &   60 \\
0503a &  5320.005 &  104 &  &   0503i2a & 5320.030 &   93 \\
0503b &  5320.143 &  138 &  &   0503i2b & 5320.162 &   92 \\
0503c &  5320.309 &  155 &  &   0503i2c & 5320.323 &  139 \\
0504a &  5320.995 &  165 &  &   0504i2a & 5321.012 &  109 \\
0504b &  5321.160 &  167 &  &   0504i2b & 5321.172 &  157 \\
0504c &  5321.317 &  141 &  &   0504i2c & 5321.326 &   79 \\
0505a &  5322.009 &  104 &  &   0505i2a & 5322.027 &   96 \\
0505b &  5322.169 &  149 &  &   0505i2b & 5322.163 &  131 \\
0505c &  5322.294 &   84 &  &   0505i2c & 5322.308 &   65 \\
\hline
\end{tabular}
\end{center}
Columns (1)--(3) are for the normal spectroscopic observations,
while columns (4)--(6) are for the special observations with I$_{2}$ cell.
The first 4 characters of the spectrum code denote the observed date
(e.g., ``0427'' means April 27), while the last character indicates
when the observation was done in a night: `a' $\cdots$ early night,
`b' $\cdots$ around mid-night, and `c' $\cdots$ late night (before dawn).
The mean signal-to-noise ratios at Columns (3) and (6) were derived
for the 5030--6260~\AA\ region data (corresponding to one of the three 
mosaicked CCDs). Otherwise, the same as described in the caption of table~1. 
\end{minipage}
\end{table*}

\setcounter{table}{2}
\begin{table*}
\begin{minipage}{180mm}
\scriptsize
\caption{Adopted spectral lines and the results of measurements.}
\begin{center}
\begin{tabular}{ccccccccccc}\hline
line code& species & $\lambda$ & $\chi_{\rm low}$ & $g_{\rm eff}^{\rm L}$ & $K$ & $n$ & 
$\langle \log W \rangle$ & $\sigma_{\log W}$ & $\langle \log v_{\rm M} \rangle$ &
$\sigma_{v}$ \\ 
(1) & (2) & (3) & (4) & (5) & (6) & (7) & (8) & (9) & (10) & (11) \\
\hline
\multicolumn{11}{c}{[orange region lines]}\\
2600\_6003010 &Fe~{\sc i} & 6003.010 & 3.882&  1.250 &  $-$2.63 & 71 & 1.9818& 0.0088&  0.6573& 0.0131\\
2800\_6007306 &Ni~{\sc i} & 6007.306 & 1.676&  1.000 &  $-$6.07 & 71 & 1.3721& 0.0148&  0.6258& 0.0313\\
2600\_6027050 &Fe~{\sc i} & 6027.050 & 4.076&  1.100 &  $-$2.39 & 71 & 1.8234& 0.0048&  0.6247& 0.0196\\
2600\_6065482 &Fe~{\sc i} & 6065.482 & 2.608&  $\cdots$ &  $-$3.53 & 71 & 2.1431& 0.0037&  0.6300& 0.0130\\
2600\_6082708 &Fe~{\sc i} & 6082.708 & 2.223&  2.000 &  $-$5.84 & 71 & 1.6048& 0.0132&  0.6472& 0.0270\\
2601\_6084111 &Fe~{\sc ii} & 6084.111 & 3.199&  0.778 &   +4.12 & 71 & 1.1283& 0.0162&  0.6386& 0.0272\\
2600\_6093666 &Fe~{\sc i} & 6093.666 & 4.607&  0.333 &  $-$3.49 & 71 & 1.4826& 0.0079&  0.6297& 0.0164\\
2600\_6094364 &Fe~{\sc i} & 6094.364 & 4.652& $-$0.250 &  $-$4.04 & 71 & 1.2888& 0.0126&  0.6512& 0.0232\\
2600\_6096662 &Fe~{\sc i} & 6096.662 & 3.984&  1.500 &  $-$3.85 & 71 & 1.5954& 0.0080&  0.6513& 0.0185\\
2600\_6098280 &Fe~{\sc i} & 6098.280 & 4.558&  1.667 &  $-$4.38 & 71 & 1.2127& 0.0206&  0.6554& 0.0314\\
2800\_6108107 &Ni~{\sc i} & 6108.107 & 1.676&  1.083 &  $-$3.05 & 71 & 1.8121& 0.0037&  0.6259& 0.0139\\
2800\_6111066 &Ni~{\sc i} & 6111.066 & 4.088&  1.250 &  $-$2.36 & 71 & 1.4680& 0.0084&  0.6401& 0.0132\\
1400\_6125021 &Si~{\sc i} & 6125.021 & 5.613& $\cdots$ &   +0.22 & 71 & 1.4018& 0.0118&  0.6620& 0.0214\\
2200\_6126217 &Ti~{\sc i} & 6126.217 & 1.067&  1.250 &  $-$9.39 & 71 & 1.4835& 0.0148&  0.6364& 0.0206\\
2600\_6127909 &Fe~{\sc i} & 6127.909 & 4.143&  0.375 &  $-$2.98 & 71 & 1.6849& 0.0062&  0.6160& 0.0163\\
2800\_6130130 &Ni~{\sc i} & 6130.130 & 4.266&  0.500 &  $-$2.68 & 71 & 1.2712& 0.0175&  0.6446& 0.0320\\
2400\_6135734 &Cr~{\sc i} & 6135.734 & 4.824&  1.333 &  $-$4.34 & 71 & 1.1978& 0.0472&  0.7165& 0.0350\\
1400\_6142483 &Si~{\sc i} & 6142.483 & 5.619& $\cdots$ &   +0.30 & 71 & 1.4457& 0.0123&  0.6565& 0.0215\\
1400\_6145016 &Si~{\sc i} & 6145.016 & 5.616& $\cdots$ &   +0.26 & 71 & 1.5009& 0.0124&  0.6420& 0.0259\\
2601\_6149258 &Fe~{\sc ii} & 6149.258 & 3.889&  1.333 &   +4.22 & 71 & 1.3910& 0.0108&  0.6244& 0.0245\\
2600\_6151617 &Fe~{\sc i} & 6151.617 & 2.176&  1.833 &  $-$4.43 & 71 & 1.7515& 0.0049&  0.6365& 0.0183\\
1100\_6154226 &Na~{\sc i} & 6154.226 & 2.102&  1.333 &  $-$5.18 & 71 & 1.5390& 0.0172&  0.6731& 0.0269\\
2600\_6157725 &Fe~{\sc i} & 6157.725 & 4.076&  1.250 &  $-$2.47 & 71 & 1.8003& 0.0039&  0.6265& 0.0172\\
2600\_6159368 &Fe~{\sc i} & 6159.368 & 4.607&  1.750 &  $-$4.57 & 71 & 1.0713& 0.0152&  0.6301& 0.0292\\
1100\_6160747 &Na~{\sc i} & 6160.747 & 2.104&  1.167 &  $-$4.29 & 71 & 1.7167& 0.0078&  0.6347& 0.0198\\
2000\_6161297 &Ca~{\sc i} & 6161.297 & 2.523&  1.333 &  $-$3.72 & 71 & 1.8390& 0.0050&  0.6234& 0.0151\\
2600\_6165361 &Fe~{\sc i} & 6165.361 & 4.143&  1.000 &  $-$3.22 & 71 & 1.6451& 0.0056&  0.6111& 0.0206\\
2000\_6166439 &Ca~{\sc i} & 6166.439 & 2.521&  0.500 &  $-$3.42 & 71 & 1.9036& 0.0045&  0.6171& 0.0162\\
2000\_6169042 &Ca~{\sc i} & 6169.042 & 2.523&  1.000 &  $-$3.24 & 71 & 2.0260& 0.0063&  0.6115& 0.0178\\
2000\_6169563 &Ca~{\sc i} & 6169.563 & 2.526&  1.167 &  $-$3.43 & 71 & 2.1323& 0.0054&  0.6061& 0.0165\\
2600\_6173341 &Fe~{\sc i} & 6173.341 & 2.223&  2.500 &  $-$3.37 & 71 & 1.9005& 0.0070&  0.6552& 0.0172\\
2800\_6175360 &Ni~{\sc i} & 6175.360 & 4.089&  1.250 &  $-$1.85 & 71 & 1.6508& 0.0037&  0.6364& 0.0155\\
2800\_6176807 &Ni~{\sc i} & 6176.807 & 4.088&  1.100 &  $-$1.43 & 71 & 1.7696& 0.0054&  0.6504& 0.0153\\
2800\_6177236 &Ni~{\sc i} & 6177.236 & 1.826&  0.500 &  $-$6.72 & 71 & 1.1185& 0.0240&  0.6141& 0.0288\\
2600\_6180203 &Fe~{\sc i} & 6180.203 & 2.727&  0.625 &  $-$3.76 & 71 & 1.7731& 0.0051&  0.6308& 0.0162\\
2800\_6186709 &Ni~{\sc i} & 6186.709 & 4.105&  1.208 &  $-$2.59 & 71 & 1.3816& 0.0112&  0.6296& 0.0203\\
2600\_6187987 &Fe~{\sc i} & 6187.987 & 3.943&  1.500 &  $-$3.34 & 71 & 1.7008& 0.0043&  0.6507& 0.0144\\
2600\_6200314 &Fe~{\sc i} & 6200.314 & 2.608& $\cdots$ &  $-$3.03 & 71 & 1.9155& 0.0062&  0.6452& 0.0216\\
2800\_6204600 &Ni~{\sc i} & 6204.600 & 4.088&  1.300 &  $-$2.79 & 71 & 1.2483& 0.0144&  0.6505& 0.0281\\
2600\_6213429 &Fe~{\sc i} & 6213.429 & 2.223&  2.000 &  $-$3.25 & 71 & 1.9956& 0.0075&  0.6424& 0.0171\\
2600\_6219279 &Fe~{\sc i} & 6219.279 & 2.198&  1.667 &  $-$3.30 & 71 & 2.0338& 0.0044&  0.6444& 0.0170\\
2800\_6223981 &Ni~{\sc i} & 6223.981 & 4.105&  1.000 &  $-$2.70 & 71 & 1.3646& 0.0108&  0.6377& 0.0175\\
2600\_6226730 &Fe~{\sc i} & 6226.730 & 3.883&  1.375 &  $-$4.61 & 71 & 1.4728& 0.0074&  0.6425& 0.0162\\
2600\_6229225 &Fe~{\sc i} & 6229.225 & 2.845&  1.000 &  $-$5.23 & 71 & 1.5918& 0.0054&  0.6179& 0.0190\\
2800\_6230090 &Ni~{\sc i} & 6230.090 & 4.105&  0.667 &  $-$2.98 & 71 & 1.1926& 0.0214&  0.6409& 0.0269\\
2600\_6232639 &Fe~{\sc i} & 6232.639 & 3.654&  2.000 &  $-$2.79 & 71 & 1.9912& 0.0048&  0.6630& 0.0144\\
2601\_6238392 &Fe~{\sc ii} & 6238.392 & 3.889&  1.467 &   +4.53 & 71 & 1.3254& 0.0147&  0.5523& 0.0226\\
2600\_6240645 &Fe~{\sc i} & 6240.645 & 2.223&  1.000 &  $-$4.38 & 71 & 1.7451& 0.0096&  0.6284& 0.0163\\
2101\_6245637 &Sc~{\sc ii} & 6245.637 & 1.507&  1.167 &   +0.24 & 71 & 1.3727& 0.0115&  0.6185& 0.0198\\
2600\_6246317 &Fe~{\sc i} & 6246.317 & 3.602&  1.583 &  $-$3.37 & 71 & 2.1710& 0.0051&  0.6633& 0.0118\\
2601\_6247557 &Fe~{\sc ii} & 6247.557 & 3.892&  1.100 &   +3.41 & 71 & 1.6157& 0.0065&  0.6298& 0.0167\\
2300\_6251827 &V ~{\sc i} & 6251.827 & 0.287&  1.587 & $-$11.79 & 71 & 1.3705& 0.0194&  0.6758& 0.0205\\
2600\_6252554 &Fe~{\sc i} & 6252.554 & 2.404&  1.083 &  $-$3.71 & 71 & 2.1579& 0.0045&  0.6305& 0.0126\\
2600\_6253829 &Fe~{\sc i} & 6253.829 & 4.733&  1.125 &  $-$4.11 & 71 & 1.2656& 0.0766&  0.6321& 0.0409\\
2200\_6258104 &Ti~{\sc i} & 6258.104 & 1.443&  1.000 &  $-$5.18 & 71 & 1.7882& 0.0073&  0.6306& 0.0258\\
\hline
\end{tabular}
\end{center}
\end{minipage}
\end{table*}

\setcounter{table}{2}
\begin{table*}
\begin{minipage}{180mm}
\scriptsize
\caption{(Continued.)}
\begin{center}
\begin{tabular}{ccccccccccc}\hline
line code& species & $\lambda$ & $\chi_{\rm low}$ & $g_{\rm eff}^{\rm L}$ & $K$ & $n$ & 
$\langle \log W \rangle$ & $\sigma_{\log W}$ & $\langle \log v_{\rm M} \rangle$ &
$\sigma_{v}$ \\ 
(1) & (2) & (3) & (4) & (5) & (6) & (7) & (8) & (9) & (10) & (11) \\
\hline
\multicolumn{11}{c}{[near IR region lines]}\\
2600\_7568894 &Fe~{\sc i} & 7568.894 & 4.283&  1.500 &  $-$2.48 & 37 & 1.9351& 0.0045&  0.6657& 0.0156\\
2600\_7582147 &Fe~{\sc i} & 7582.147 & 4.956&  1.500 &  $-$3.79 & 37 & 1.0009& 0.0334&  0.6326& 0.0301\\
2600\_7583787 &Fe~{\sc i} & 7583.787 & 3.018&  0.833 &  $-$2.94 & 37 & 1.9632& 0.0063&  0.6284& 0.0157\\
2600\_7586014 &Fe~{\sc i} & 7586.014 & 4.312&  1.300 &  $-$2.91 & 37 & 2.1788& 0.0044&  0.6754& 0.0129\\
2600\_7748274 &Fe~{\sc i} & 7748.274 & 2.949&  1.100 &  $-$3.00 & 59 & 2.0487& 0.0047&  0.6356& 0.0181\\
2600\_7751137 &Fe~{\sc i} & 7751.137 & 4.991&  1.200 &  $-$2.51 & 59 & 1.6664& 0.0081&  0.6583& 0.0227\\
0800\_7774166 &O ~{\sc i} & 7774.166 & 9.146&  1.917 &   +6.59 & 59 & 1.6324& 0.0135&  0.6735& 0.0188\\
2600\_7780552 &Fe~{\sc i} & 7780.552 & 4.473&  0.833 &  $-$2.48 & 59 & 2.1380& 0.0034&  0.6659& 0.0115\\
2600\_7802473 &Fe~{\sc i} & 7802.473 & 5.085&  1.500 &  $-$3.54 & 59 & 1.2205& 0.0527&  0.6674& 0.0416\\
2600\_7807952 &Fe~{\sc i} & 7807.952 & 4.991&  1.400 &  $-$2.08 & 59 & 1.7896& 0.0073&  0.6414& 0.0132\\
2600\_7844555 &Fe~{\sc i} & 7844.555 & 4.835&  1.500 &  $-$4.25 & 59 & 1.0540& 0.0372&  0.6423& 0.0504\\
2600\_7941094 &Fe~{\sc i} & 7941.094 & 3.274&  0.500 &  $-$4.45 & 59 & 1.6360& 0.0208&  0.6119& 0.0346\\
2600\_8047615 &Fe~{\sc i} & 8047.615 & 0.859&  1.500 &  $-$5.35 & 59 & 1.8338& 0.0088&  0.6434& 0.0229\\
2600\_8207745 &Fe~{\sc i} & 8207.745 & 4.446&  0.667 &  $-$2.51 & 59 & 1.8583& 0.0102&  0.6386& 0.0210\\
2600\_8248120 &Fe~{\sc i} & 8248.120 & 4.371&  1.250 &  $-$2.66 & 59 & 1.8103& 0.0275&  0.6560& 0.0408\\
2600\_8365633 &Fe~{\sc i} & 8365.633 & 3.251&  1.333 &  $-$3.22 & 59 & 1.8832& 0.0112&  0.6524& 0.0301\\
2600\_8468404 &Fe~{\sc i} & 8468.404 & 2.223&  2.500 &  $-$4.14 & 59 & 2.2328& 0.0114&  0.6835& 0.0177\\
2600\_8471739 &Fe~{\sc i} & 8471.739 & 4.956&  1.500 &  $-$2.98 & 59 & 1.5471& 0.0213&  0.6498& 0.0331\\
2600\_8514069 &Fe~{\sc i} & 8514.069 & 2.198&  1.833 &  $-$4.00 & 59 & 2.2085& 0.0108&  0.6892& 0.0215\\
2600\_8515110 &Fe~{\sc i} & 8515.110 & 3.018&  0.750 &  $-$3.03 & 59 & 2.0104& 0.0075&  0.6713& 0.0232\\
2600\_8571802 &Fe~{\sc i} & 8571.802 & 5.009&  2.000 &  $-$2.95 & 59 & 1.4714& 0.0331&  0.6878& 0.0401\\
2600\_8582257 &Fe~{\sc i} & 8582.257 & 2.990&  1.050 &  $-$3.22 & 59 & 1.9220& 0.0089&  0.6270& 0.0191\\
2600\_8592945 &Fe~{\sc i} & 8592.945 & 4.956&  1.375 &  $-$2.41 & 59 & 1.7204& 0.0138&  0.7142& 0.0250\\
2600\_8598825 &Fe~{\sc i} & 8598.825 & 4.386&  1.300 &  $-$2.94 & 59 & 1.7682& 0.0131&  0.6628& 0.0250\\
2600\_8611795 &Fe~{\sc i} & 8611.795 & 2.845&  1.500 &  $-$3.09 & 59 & 2.0729& 0.0078&  0.6624& 0.0262\\
2600\_8613935 &Fe~{\sc i} & 8613.935 & 4.988&  1.833 &  $-$3.08 & 59 & 1.4839& 0.0277&  0.6887& 0.0466\\
2600\_8616276 &Fe~{\sc i} & 8616.276 & 4.913&  1.100 &  $-$2.67 & 59 & 1.6474& 0.0217&  0.6551& 0.0257\\
2600\_8621598 &Fe~{\sc i} & 8621.598 & 2.949&  1.200 &  $-$3.25 & 59 & 1.8994& 0.0079&  0.6485& 0.0251\\
2600\_8674743 &Fe~{\sc i} & 8674.743 & 2.832&  1.500 &  $-$3.21 & 59 & 2.1036& 0.0077&  0.6509& 0.0141\\
2600\_8688621 &Fe~{\sc i} & 8688.621 & 2.176&  1.667 &  $-$5.46 & 59 & 2.5159& 0.0049&  0.6822& 0.0122\\
2600\_8699446 &Fe~{\sc i} & 8699.446 & 4.956&  1.125 &  $-$2.07 & 59 & 1.8433& 0.0115&  0.6481& 0.0181\\
2600\_8729148 &Fe~{\sc i} & 8729.148 & 3.415&  0.750 &  $-$5.59 & 59 & 1.4073& 0.0337&  0.6921& 0.0515\\
2600\_8747423 &Fe~{\sc i} & 8747.423 & 3.018&  1.500 &  $-$6.68 & 59 & 1.3310& 0.0504&  0.6950& 0.0686\\
2600\_8757182 &Fe~{\sc i} & 8757.182 & 2.845&  1.500 &  $-$3.10 & 59 & 2.0431& 0.0071&  0.6592& 0.0232\\
2600\_8763962 &Fe~{\sc i} & 8763.962 & 4.652&  0.667 &  $-$2.19 & 59 & 2.0614& 0.0075&  0.6822& 0.0187\\
2600\_8784434 &Fe~{\sc i} & 8784.434 & 4.956&  1.500 &  $-$3.24 & 59 & 1.4545& 0.0338&  0.6661& 0.0381\\
2600\_8796478 &Fe~{\sc i} & 8796.478 & 4.956&  1.000 &  $-$3.28 & 59 & 1.4055& 0.0411&  0.6670& 0.0380\\
2600\_8804623 &Fe~{\sc i} & 8804.623 & 2.279& $\cdots$ &  $-$4.25 & 59 & 1.8298& 0.0122&  0.6411& 0.0194\\
2600\_8824216 &Fe~{\sc i} & 8824.216 & 2.198&  1.500 &  $-$4.99 & 37 & 2.3902& 0.0052&  0.6690& 0.0161\\
2600\_8838423 &Fe~{\sc i} & 8838.423 & 2.858&  1.500 &  $-$3.07 & 37 & 2.0567& 0.0094&  0.6582& 0.0185\\
2600\_8846736 &Fe~{\sc i} & 8846.736 & 5.009&  0.750 &  $-$2.32 & 37 & 1.7117& 0.0208&  0.6687& 0.0248\\
2600\_8868431 &Fe~{\sc i} & 8868.431 & 3.018&  0.833 &  $-$4.31 & 37 & 1.7413& 0.0158&  0.6620& 0.0271\\
2600\_8876022 &Fe~{\sc i} & 8876.022 & 5.020&  0.000 &  $-$2.97 & 37 & 1.5066& 0.0185&  0.6469& 0.0342\\
2600\_8920018 &Fe~{\sc i} & 8920.018 & 5.064&  1.000 &  $-$2.23 & 37 & 1.7627& 0.0311&  0.6633& 0.0342\\
\hline
\end{tabular}
\end{center}
Column (1) --- 12-character line code indicating the species and wavelength 
(according to the same definition as adopted by Takeda \& UeNo 2019).
Column (2) --- element species. 
Column (3) --- line wavelength (in \AA).
Column (4) --- lower excitation potential (in eV).
Column (5) --- effective Land\'{e} factor computed from $L$, $S$, and $J$ values
of the upper and lower levels (kept blank when the relevant term data 
are not available). 
Column (6) --- Temperature-sensitivity index (${\rm d}\log W/{\rm d}\log T$)
computed by equation~(7).
Column (7) --- number of available spectra.
Column (8) --- Average of $\log W$ ($W$ is the equivalent width in m\AA).
Column (9) --- Standard deviation of $\log W$ around the mean.
Column (10) --- Average of $\log v_{\rm M}$ ($v_{\rm M}$ is the macrobroadening
velocity in km~s$^{-1}$).
Column (11) --- Standard deviation of $\log v_{\rm M}$ around the mean.\\
Note: Regarding the derivation of average-related results presented in Columns (8)--(11),  
those data points showing appreciable deviations, which were judged by Chauvenet's 
criterion (Taylor 1997), were discarded (the rejected data are distinguished by 
the negative sign in tableE4.dat).

\end{minipage}
\end{table*}

\newpage

\setcounter{figure}{0}
\begin{figure*}[p]
  \begin{center}
    \FigureFile(100mm,160mm){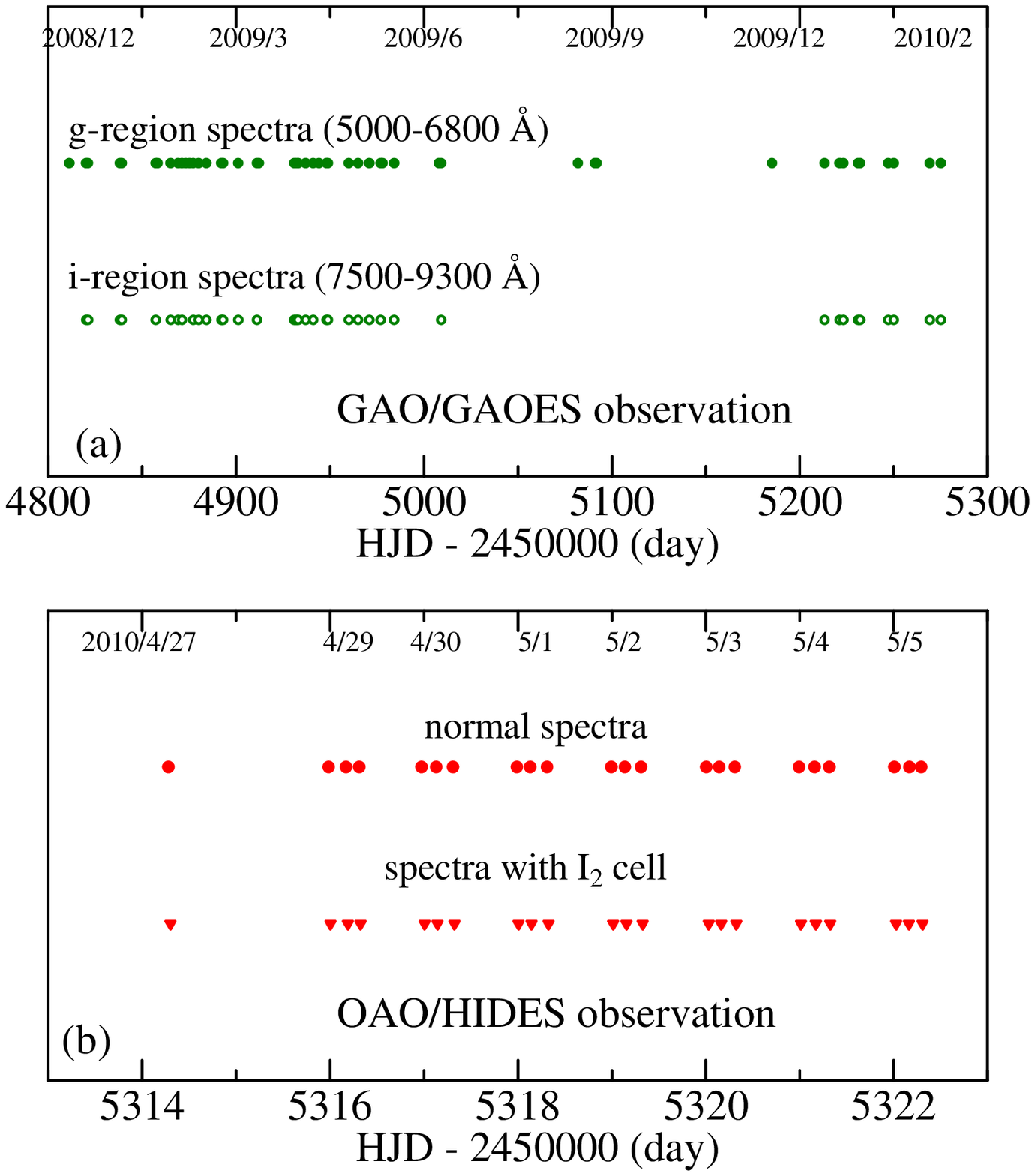}
  \end{center}
\caption{
Graphical representation of the observed dates for the spectra used in this study. 
(a) GAO/GAOES observations from 2008 December to 2010 march (g-region spectra 
and i-region spectra). 
(b) OAO/HIDES observations in 2010 late April and early May (normal spectra
and spectra with I$_{2}$ cell). 
}
\end{figure*}

\setcounter{figure}{1}
\begin{figure*}[p]
  \begin{center}
    \FigureFile(120mm,150mm){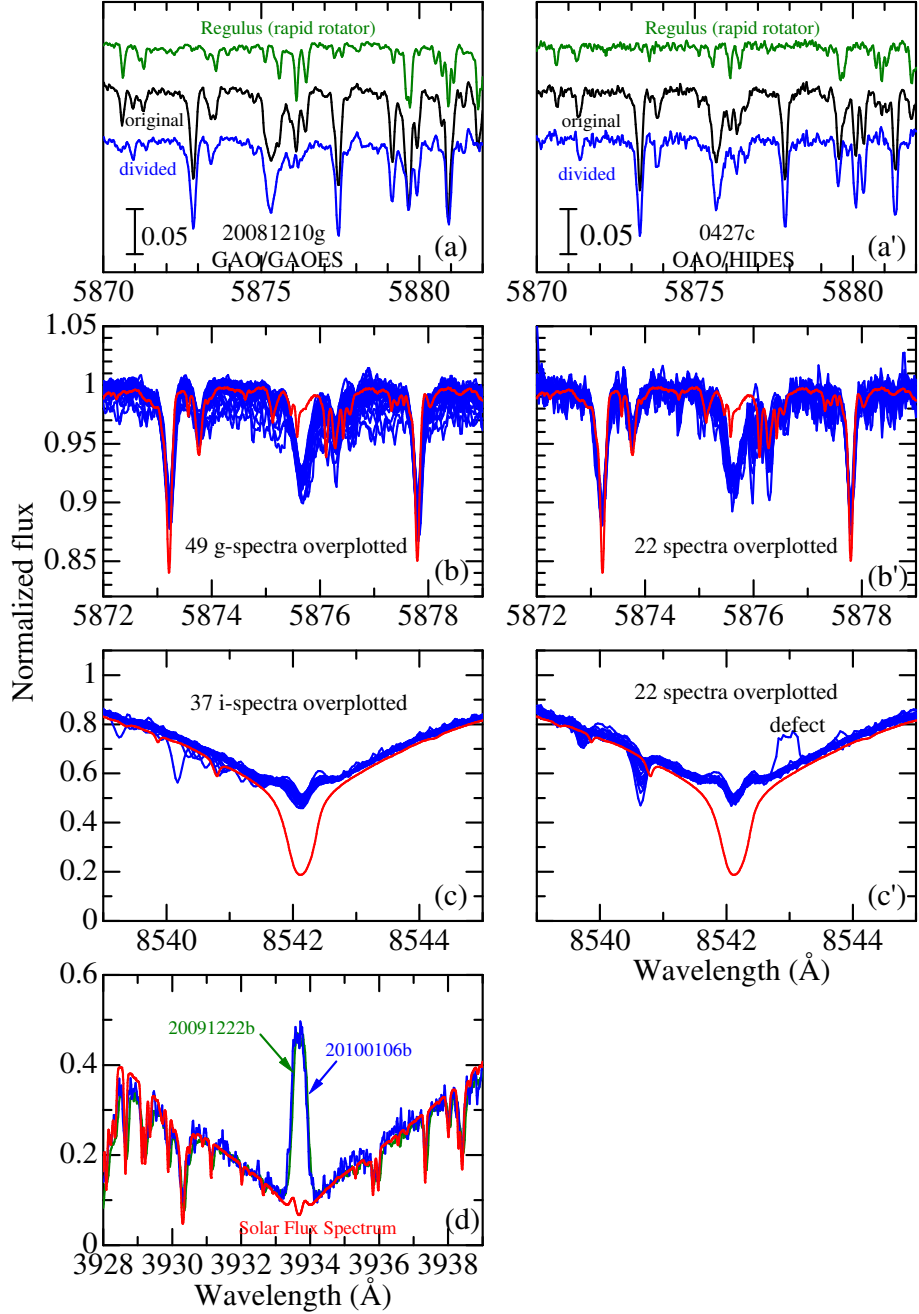}
  \end{center}
\caption{
Display of spectral features for three activity-sensitive lines. 
The left (a--d) and right (a$'$--c$'$) panels correspond to GAOES and 
HIDES spectra, respectively.
(a),(a$'$) $\cdots$ examples of how the telluric lines are removed 
in the neighborhood of He~{\sc i} 5876 line, where the spectra 
are depicted in the raw (uncorrected) wavelength scale.
(b),(b$'$) $\cdots$ spectra of 5872--5879~\AA\ region comprising 
the He~{\sc i} 5876 line overplotted (wavelength scale adjusted to the
laboratory frame).
(c),($c'$) $\cdots$ spectra of 8539--8545~\AA\ region comprising
the Ca~{\sc ii} 8542 line overplotted (wavelength scale adjusted to the
laboratory frame).
(d) $\cdots$ spectra of 3928--3939~\AA\ region comprising the Ca~{\sc ii} 3934 
line (only two GAOES spectra obtained on 2009 December 22 and
2010 January 6 overplotted). In each panel, the relevant solar flux spectra
(taken from Kurucz et al. 1984) are shown by red lines (note that telluric lines
are not removed in these spectra of the Sun). 
}
\end{figure*}

\setcounter{figure}{2}
\begin{figure*}[p]
  \begin{center}
    \FigureFile(130mm,160mm){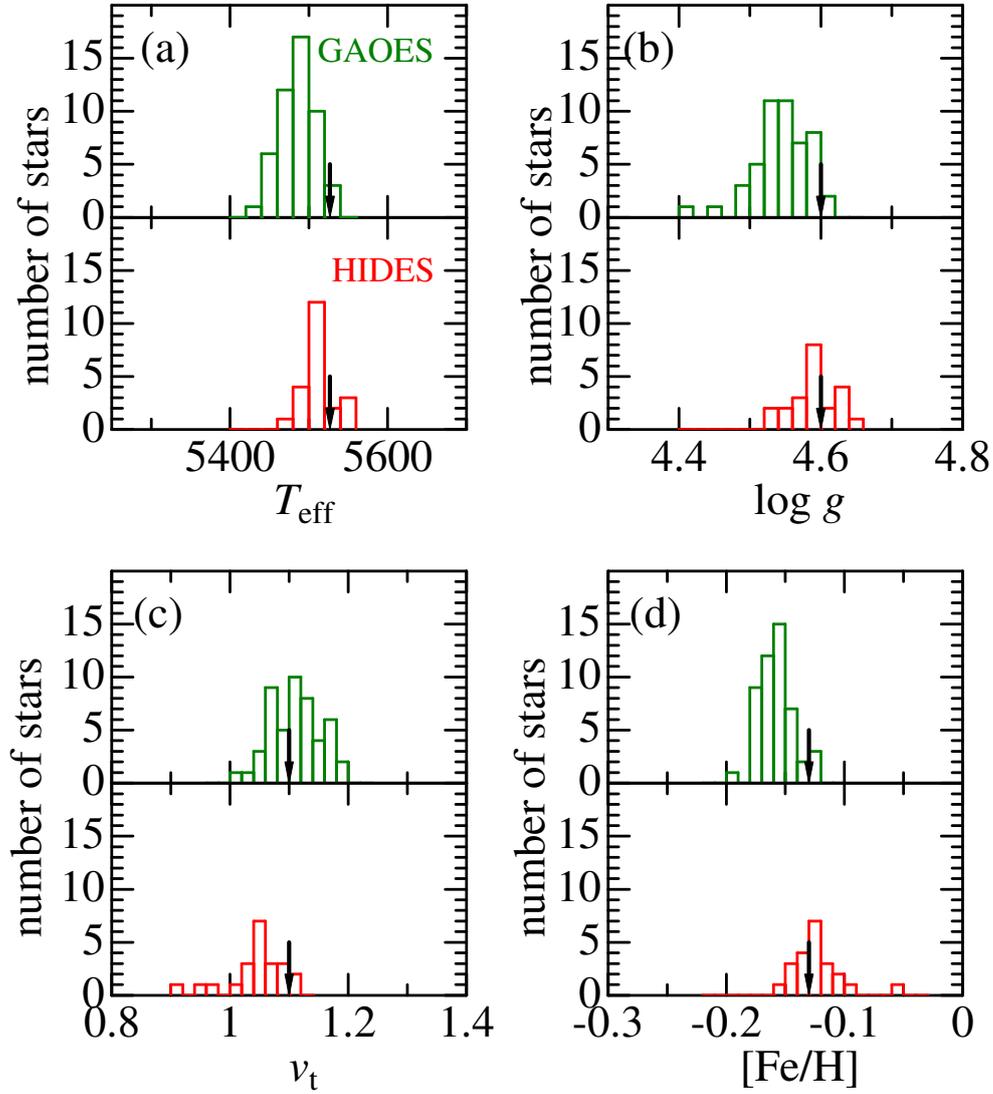}
  \end{center}
\caption{
Histograms showing the distributions of spectroscopically determined
atmospheric parameters: (a) $T_{\rm eff}$ (in K), (b) $\log g$ (in dex, 
where $g$ is in unit of cm~s$^{-2}$), (c) $v_{\rm t}$ (in km~s$^{-1}$), 
and (d) [Fe/H] (in dex). The results based on the GAOES spectra
(upper panel) and HIDES spectra (lower panel) are separately shown. 
The downward arrows indicate the positions of the standard values 
(5527~K, 4.60~dex, 1.10~km~s$^{-1}$, and $-0.13$~dex) determined by Takeda et al. (2005). 
}
\end{figure*}

\setcounter{figure}{3}
\begin{figure*}[p]
  \begin{center}
    \FigureFile(110mm,200mm){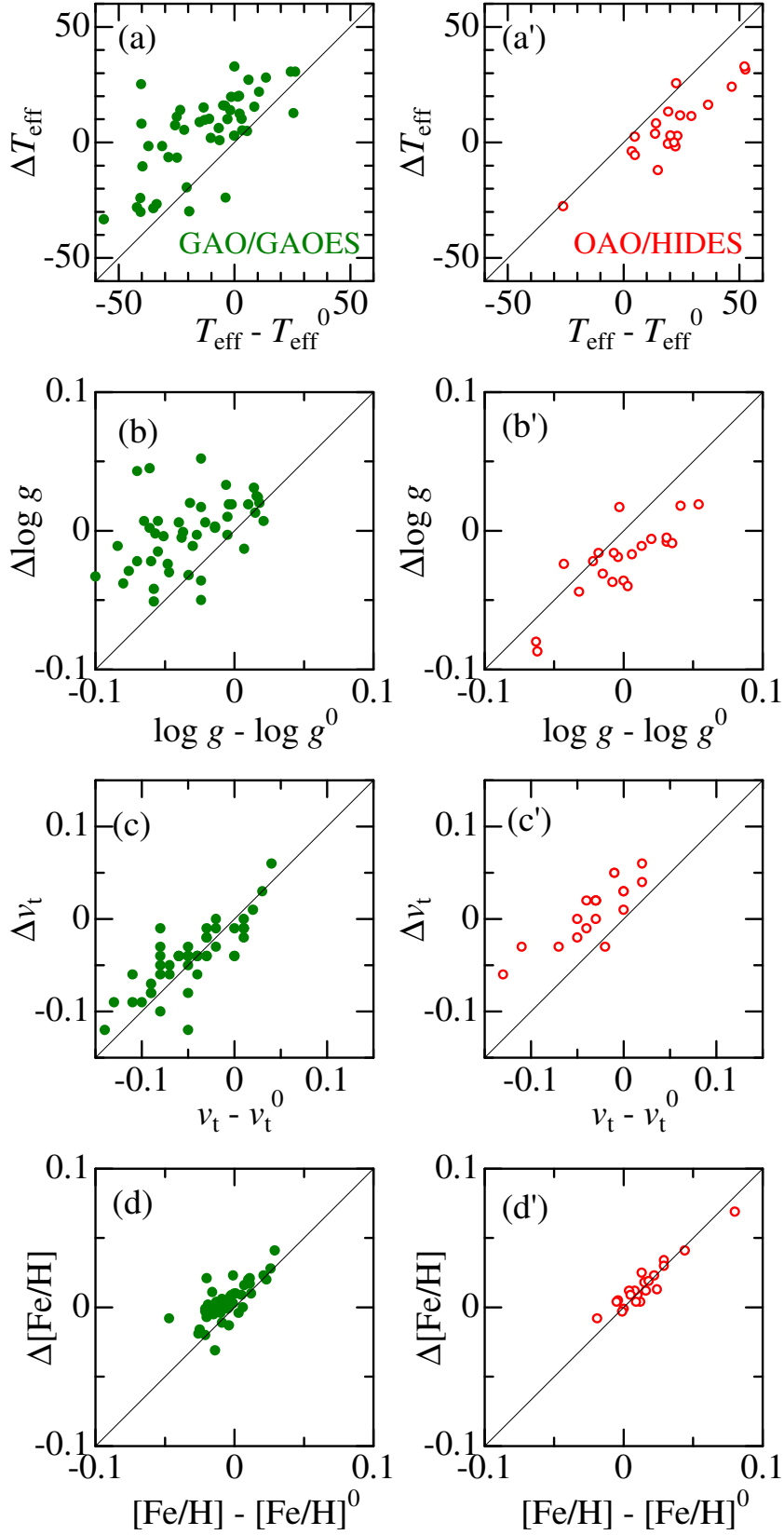}
  \end{center}
\caption{
Comparison of two kinds of differential atmospheric parameters 
($T_{\rm eff}$, $\log g$, $v_{\rm t}$, and [Fe/H]) relative to the 
reference values (denoted with superscript ``0'') corresponding to 
the first observation: 20081210g for the GAOES results (left panels a--d)
and 0427c for the HIDES results (right panels a$'$--d$'$),
Abscissa: simple differences of the absolute parameter values derived 
by the conventional method (cf. subsection~3.1).
Ordinates: results obtained by the method of purely differential 
parameter determination (cf. subsection~3.2). 
}
\end{figure*}

\setcounter{figure}{4}
\begin{figure*}[p]
  \begin{center}
    \FigureFile(90mm,120mm){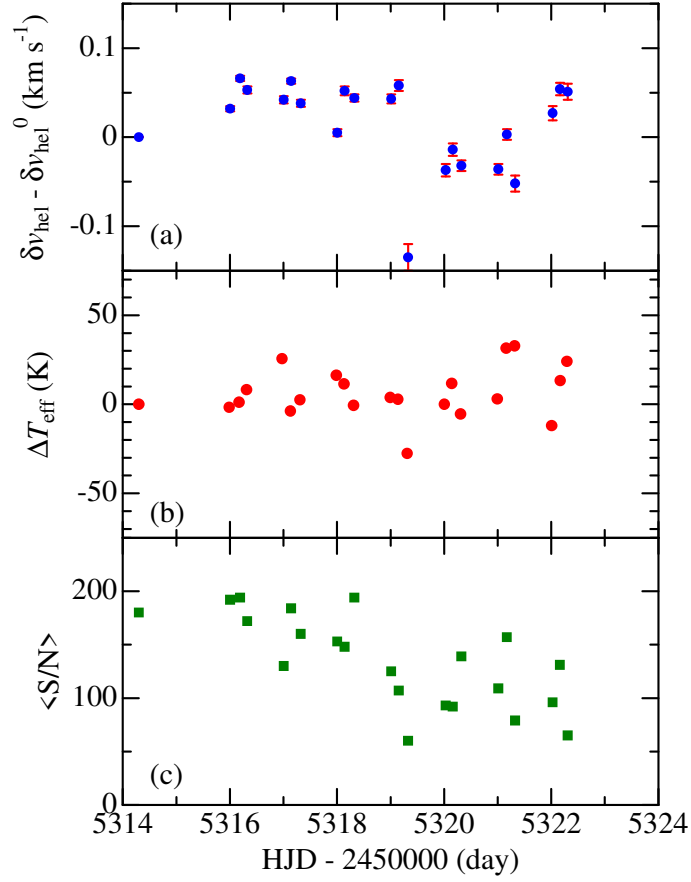}
  \end{center}
\caption{
(a) Differential heliocentric radial velocities relative to the first observation
of 0427i2c on HJD 2455314.30 (cf. tableE3.dat), which were derived by analyzing the 
22 OAO/HIDES spectra obtained with I$_{2}$ cell, plotted against the observation time.
The error bars indicate the probable errors.
(b) Variations of effective temperature (relative to that of 0427c) derived by the
differential parameter analysis (cf. subsection~3.2) plotted against the observation time.
(c) Signal-to-noise ratios of the I$_{2}$ spectra (cf. table~2) plotted against 
the observation time.
}
\end{figure*}

\setcounter{figure}{5}
\begin{figure*}[p]
  \begin{center}
    \FigureFile(85mm,100mm){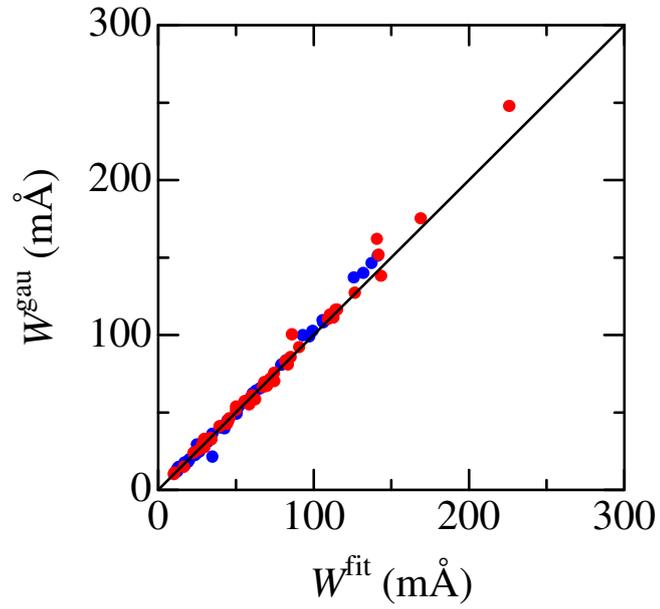}
  \end{center}
\caption{
Comparison of the equivalent widths of 99 lines derived by the spectrum-fitting 
method adopted in this study (abscissa) and those directly measured by 
the conventional Gaussian-fitting technique (ordinate). 
Blue and red symbols correspond to lines in the 6000--6260~\AA\ region
and those in the 7560--8920~\AA\ region, respectively. 
}
\end{figure*}

\setcounter{figure}{6}
\begin{figure*}[p]
  \begin{center}
    \FigureFile(115mm,160mm){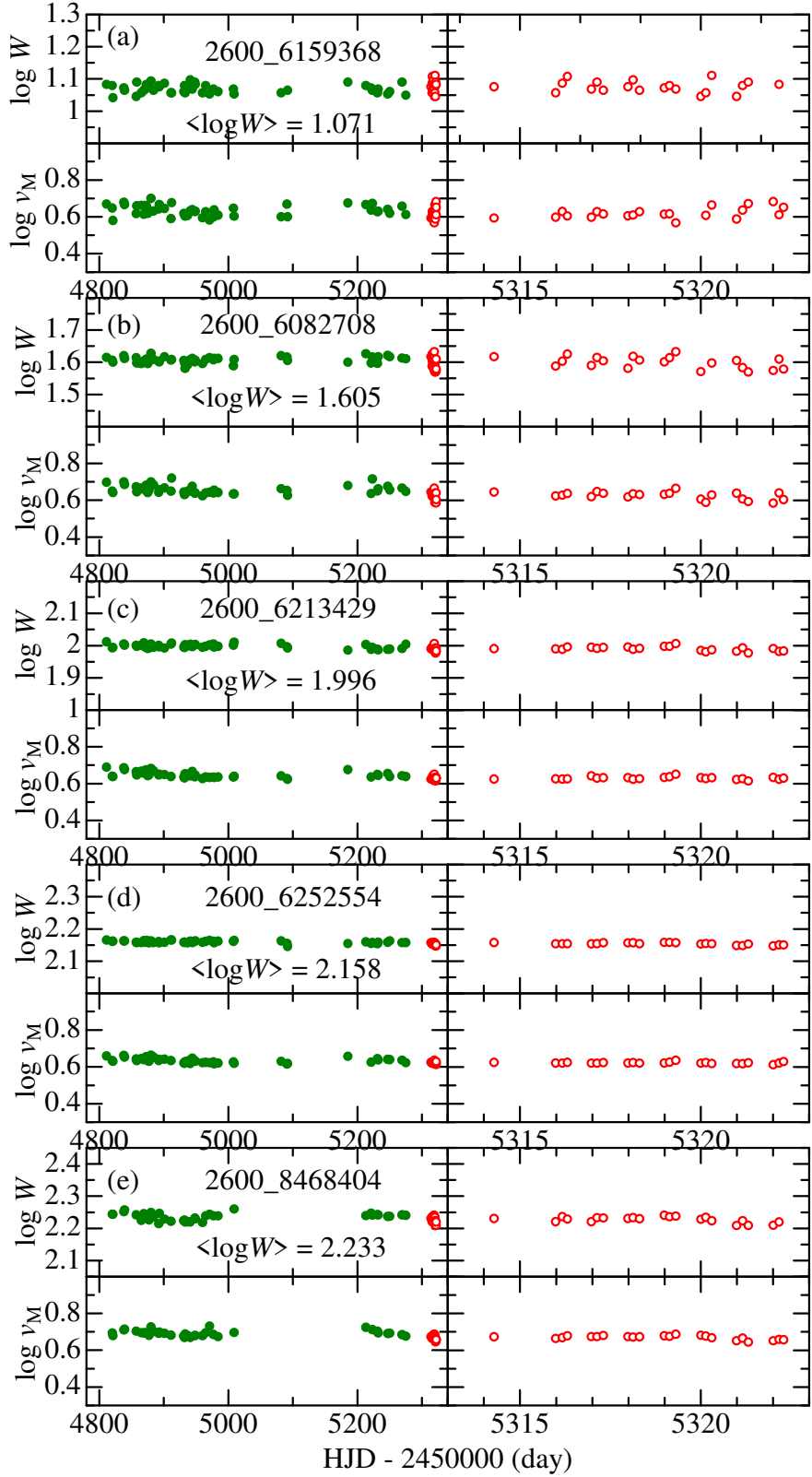}
  \end{center}
\caption{
Logarithmic equivalent widths ($W$ in m\AA) and macrobroadening velocities
($v_{\rm M}$ in km~s$^{-1}$) of representative 5 Fe~{\sc i} lines of different strengths,
plotted against the observational time.
(a) Fe~{\sc i} 6159.368 ($\langle \log W \rangle = 1.071$),
(b) Fe~{\sc i} 6082.708 ($\langle \log W \rangle = 1.605$),
(c) Fe~{\sc i} 6213.429 ($\langle \log W \rangle = 1.996$), 
(d) Fe~{\sc i} 6252.554 ($\langle \log W \rangle = 2.158$), and
(e) Fe~{\sc i} 8468.404 ($\langle \log W \rangle = 2.233$).
Filled and open symbols correspond to the results based on GAO/GAOES and OAO/HIDES
spectra, respectively. While all the data are plotted in the left panels,
in the right panels are shown only the OAO/HIDES data in the expanded scale. 
}
\end{figure*}

\setcounter{figure}{7}
\begin{figure*}[p]
  \begin{center}
    \FigureFile(130mm,180mm){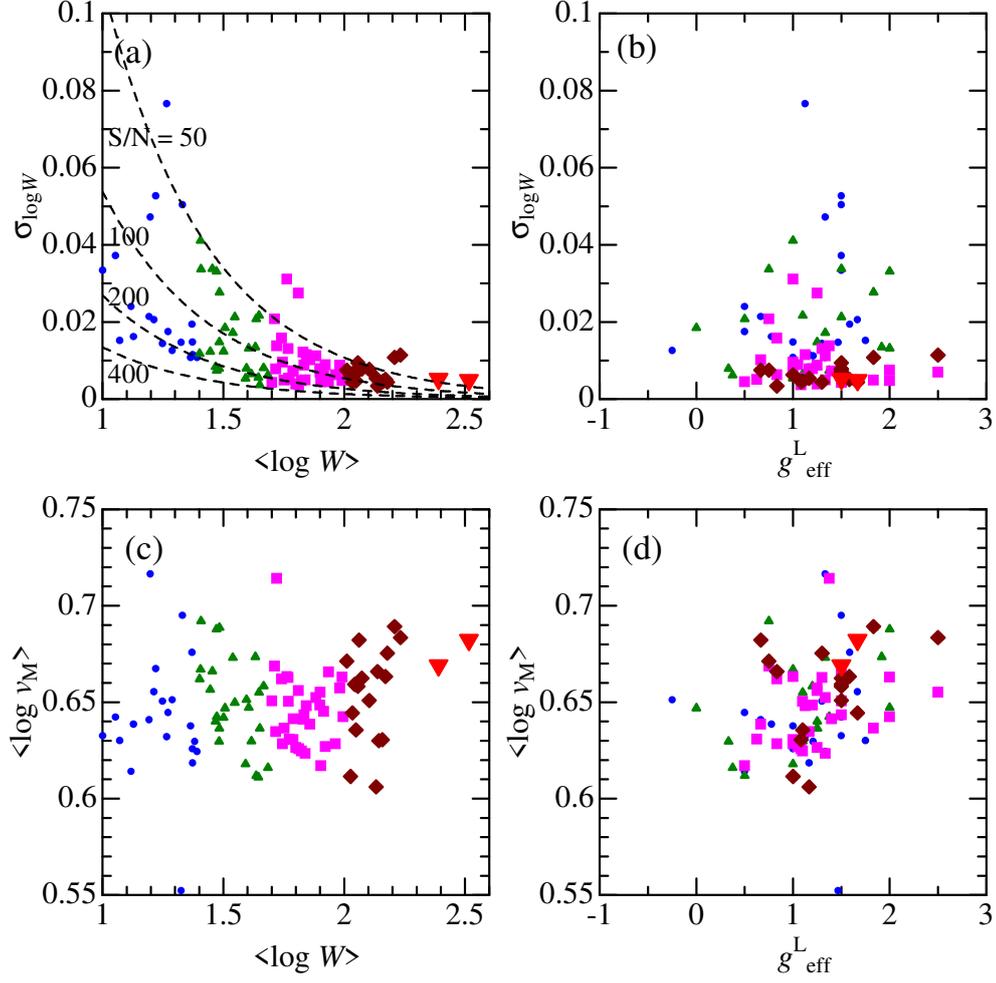}
  \end{center}
\caption{
Upper panels: standard deviations of $\log W$ plotted against
(a) $\langle \log W \rangle$ (mean of $\log W$) and
(b) $g_{\rm eff}^{\rm L}$ (effective Land\'{e} factor).
Lower panels:  mean of $\log v_{\rm M}$ ($v_{\rm M}$: macrobroadening velocity
in km~s$^{-1}$) plotted against (c) $\langle \log W \rangle$ and
(d) $g_{\rm eff}^{\rm L}$.
In panel (a) are also shown by dashed lines the expected $\delta \log W$ 
(S/N-dependent random error in $\log W$) vs. $W$ relations (cf. 
equations~(5) and (6)) for S/N = 50, 100, 200, and 400. 
Lines of different strengths classes are discriminated by 
the shape and the size (larger for stronger lines) of symbols:
circles (blue): $\langle \log W \rangle < 1.4$, 
triangles (green): $1.4 \le \langle \log W \rangle <1.7$,
squares (pink): $1.7 \le \langle \log W \rangle < 2.0$, 
diamonds (brown): $2.0 \le \langle \log W \rangle < 2.3$,
and inverse triangles (red):  2.3~$\le \langle \log W \rangle$.
}
\end{figure*}

\setcounter{figure}{8}
\begin{figure*}[p]
  \begin{center}
    \FigureFile(80mm,100mm){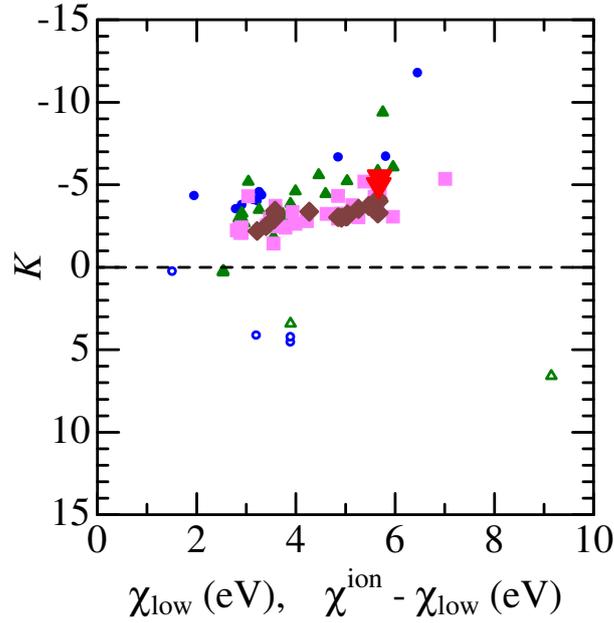}
  \end{center}
\caption{
Temperature-sensitivity parameter $K$ ($\equiv {\rm d}\log W/{\rm d}\log T$) 
computed for 99 lines by equation~(7), plotted against $\chi_{\rm low}$ 
(for major population species: O~{\sc i}, Sc~{\sc ii}, and Fe~{\sc ii}) 
or $\chi^{\rm ion} - \chi_{\rm low}$ 
(for minor population species: Na~{\sc i}, Si~{\sc i}, Ca~{\sc i}, 
Ti~{\sc i}, V~{\sc i}, Fe~{\sc i}, and Ni~{\sc i}).
The filled and open symbols correspond to minor population
and major population species, respectively.
Lines of different strengths classes (judged by the conventionally measured
equivalent widths $W^{\rm gau}$; cf. figure~6) are discriminated by 
the shape and the size (larger for stronger lines) of symbols:
circles (blue): $W^{\rm gau} <$~25~m\AA\, 
triangles (green): 25~m\AA~$\le W^{\rm gau} <$~50~m\AA\,
squares (pink): 50~m\AA~$\le W^{\rm gau} <$~100~m\AA\, 
diamonds (brown): 100~m\AA~$\le W^{\rm gau} <$~200~m\AA\,
and inverse triangles (red):  200~m\AA~$\le W^{\rm gau}$.
}
\end{figure*}

\setcounter{figure}{9}
\begin{figure*}[p]
  \begin{center}
    \FigureFile(130mm,200mm){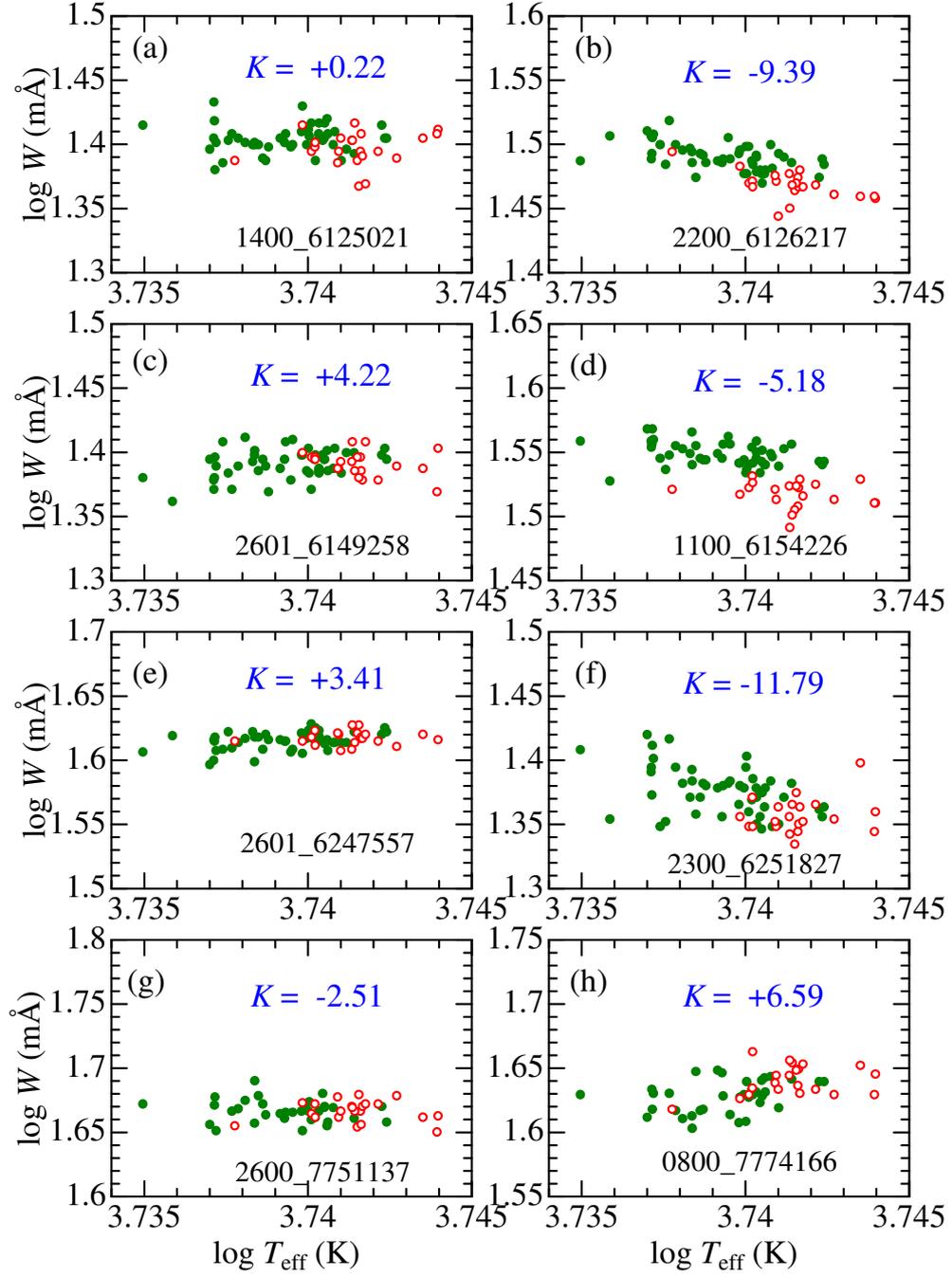}
  \end{center}
\caption{
Logarithmic equivalent widths ($W$ in m\AA) of 8 medium-strength lines
($\langle \log W \rangle \sim$~1.4--1.6) with different temperature 
sensitivity ($K$), which were measured for each of the spectra at different 
observational times, plotted against $\log T_{\rm eff}$.
(a) Si~{\sc i} 6125.021 ($K = +0.22$),
(b) Ti~{\sc i} 6126.217 ($K = -9.39$),
(c) Fe~{\sc ii} 6149.258 ($K = +4.22$), 
(d) Na~{\sc i} 6154.226 ($K = -5.18$),
(e) Fe~{\sc ii} 6247.557 ($K = +3.41$),
(f) V~{\sc i} 6251.827 ($K = -11.79$),
(g) Fe~{\sc i} 7751.137 ($K = -2.51$), and
(h) O~{\sc i} 7774.166 ($K = +6.59$).
The same meanings of the symbols as in figure~7.
}
\end{figure*}

\setcounter{figure}{10}
\begin{figure*}[p]
  \begin{center}
    \FigureFile(120mm,200mm){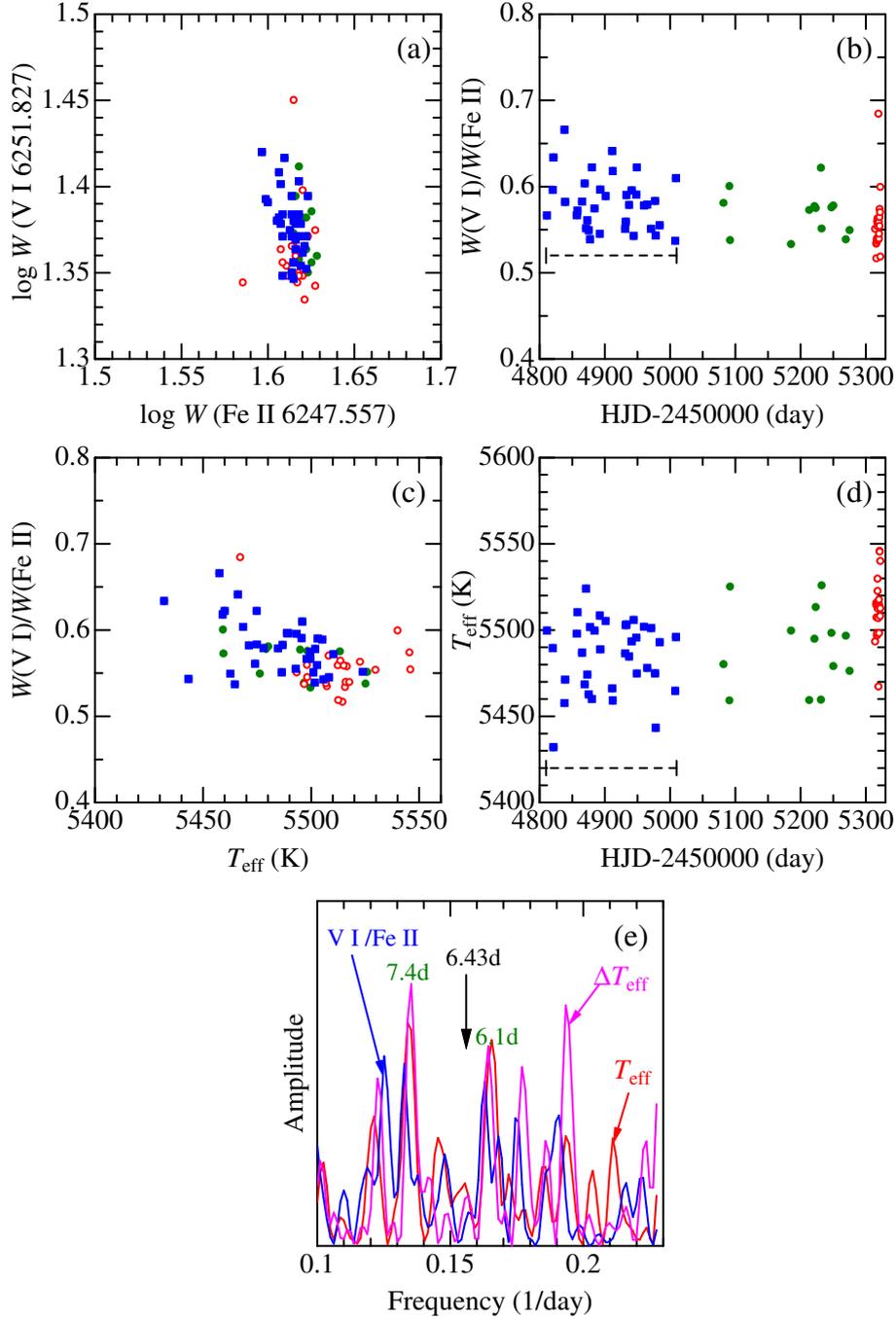}
  \end{center}
\caption{
(a) Correlation of $\log W$(V~{\sc i} 6251.827)
and $\log W$(Fe~{\sc ii} 6247.557).
(b) V~{\sc i} to Fe~{\sc ii} equivalent width ratio
plotted against the observed dates. 
(c) V~{\sc i} to Fe~{\sc ii} equivalent width ratio
plotted against $T_{\rm eff}$.
(d) $T_{\rm eff}$ vs. observed dates.
(e) Power spectra of the time-series data (corresponding to the period of 
$\sim 200$ days from HJD 2454810 to 2455010; cf. the range indicated by 
the horizontal dashed line in panels b and d) for V~{\sc i} to Fe~{\sc ii} 
equivalent width ratio (blue line), $T_{\rm eff}$ (absolute values; red line) and 
$\Delta T_{\rm eff}$ (differential values; pink line).
The meanings of the symbols in panels (a)--(d) are almost the same as in figure~7,
while those corresponding to the data used for power spectrum analysis are 
distinguished by blue filled squares.
}
\end{figure*}

\end{document}